\documentclass[10pt,
							 twocolumn,
							 superscriptaddress,
							 english,
							 prl,
							 showpacs,
							 floatfix,
							 aps
							]{revtex4-1}
\usepackage[utf8]{inputenc}
\usepackage{amsmath}
\usepackage{amssymb}
\usepackage{graphicx}
\usepackage{xspace}
\usepackage{color}
\makeatletter
\@ifundefined{textcolor}{}
{%
 \definecolor{BLACK}{gray}{0}
 \definecolor{WHITE}{gray}{1}
 \definecolor{RED}{rgb}{1,0,0}
 \definecolor{GREEN}{rgb}{0,1,0}
 \definecolor{BLUE}{rgb}{0,0,1}
 \definecolor{CYAN}{cmyk}{1,0,0,0}
 \definecolor{MAGENTA}{cmyk}{0,1,0,0}
 \definecolor{YELLOW}{cmyk}{0,0,1,0}
}

\usepackage{soul}
\usepackage{braket}
\usepackage[backref=none,
bookmarksnumbered=true,
bookmarks=true,
bookmarksopen=true,
colorlinks=true,
citecolor=blue,
linkcolor=blue,
anchorcolor=green,
urlcolor=blue,unicode=false]{hyperref}

\renewcommand{\v}[1]{\ensuremath{\mathbf{#1}}} 
 
\let\baraccent=\= 
\renewcommand{\=}[1]{\stackrel{#1}{=}} 

\newcommand{\unitspace}{~}

\newcommand{\Fig}[1]{Fig.\unitspace\ref{fig:#1}}
\newcommand{\Figure}[1]{Figure\unitspace\ref{fig:#1}}

\makeatother

\usepackage{babel}

\begin{document}

\title{Higher order topological superconductors as generators of quantum codes}

\author{Yizhi You}
\affiliation{\mbox{Princeton Center for Theoretical Science, Princeton University, Princeton NJ 08544, USA}}

\author{Daniel Litinski}
\affiliation{\mbox{Dahlem Center for Complex Quantum Systems and Fachbereich Physik, Freie Universit\"at Berlin, 14195 Berlin, Germany}}

\author{Felix von Oppen}
\affiliation{\mbox{Dahlem Center for Complex Quantum Systems and Fachbereich Physik, Freie Universit\"at Berlin, 14195 Berlin, Germany}}

\affiliation{Institute of Quantum Information and Matter, California Institute of Technology, Pasadena, California 91125, USA}

\date{\today}

\begin{abstract}
We show that interactions can drive a class of higher order topological superconductors (HOTSC) into symmetry enriched topologically ordered phases exemplified by topological quantum error correcting codes. In two dimensions, interacting HOTSC realize various topologically ordered surface and color codes. In three dimensions, interactions can drive HOTSC protected by subsystem symmetries into recently discovered fracton phases. We explicitly relate fermion parity operators underlying the gapless excitations of the HOTSC to the Wilson algebra of symmetry enriched quantum codes. Arrays of crossed Majorana wires provide an experimental platform for realizing fracton matter and for probing the quantum phase transition between HOTSC and topologically ordered phase.

\end{abstract}

\maketitle 

{\em Introduction.---}A decade of intense effort has resulted in a thorough classification and characterization of topological materials. For a refined classification of topological insulators and superconductors along with their bosonic analogs, the concept of symmetry protection has been extended to include spatial symmetries~\cite{fu2011topological,hsieh2012topological,cheng2016translational,ando2015topological,slager2013space,hong2017topological,qi2015anomalous,huang2017building,teo2013existence,song2017topological,watanabe2017structure,po2017symmetry,isobe2015theory}. In addition to fully dispersive boundary modes, topological crystalline phases admit gapped edges or surfaces with protected gapless modes at high-symmetry corners or hinges. Exemplifying a much richer bulk-boundary correspondence, this phenomenology is now termed higher-order topological 
phase~\cite{benalcazar2017quantized,benalcazar2017electric,langbehn2017reflection,song2017d,song2017d,dwivedi2018majorana} and a variety of corresponding candidate materials have been proposed \cite{schindler2018higher,benalcazar2014classification,raghu2010hidden,serra2018observation,andrei}.

An important research frontier on higher-order phases concerns the effects of strong interactions. While previous work shows that interactions can both trivialize and enrich these phases \cite{song2017topological,song2017interaction,isobe2015theory,you2018higher,2018arXiv180907325R}, most efforts to date have focused on mathematical classification rather than microscopic Hamiltonians and the ensuing phase diagrams of higher-order materials. Here, we demonstrate that strong local interactions can induce more exotic topologically ordered phases in a class of higher order topological superconductors (HOTSC).

Indeed, interactions can promote these HOTSC into phases exemplified by quantum error correcting codes with long-range entanglement and fractionalization. A connection between band theories of Majoranas and quantum codes can be traced back to Wen's mean-field description for the toric (surface) code \cite{wen2003quantum,you2012projective,hsieh2017topological}. We show that this mean-field theory typifies a HOTSC phase. A HOTSC carries a Majorana zero mode at corners protected by, e.g., reflection or rotation symmetries. Similarly, for certain boundary conditions,  
surface code patches have degenerate ground states which underlie their use as logical qubits. This degeneracy can also be thought of as originating from corners with a fermion zero mode. These corners are meeting points of edges with $e$ and $m$ condensates, with their existence enforced by symmetry. We show explicitly how mean-field theory connects symmetry-protected HOTSC and symmetry-enriched surface codes.

Similar constructions can be used to generate the recently discovered fracton codes from Majorana band models with strong interactions \cite{Vijay2016-dr,hsieh2017fractons,Slagle2017-ne,Halasz2017-ov,Hsieh2017-sc,Chamon2005-fc}. Fracton matter has been intensively explored via exactly solvable models, including quantum stabilizer codes as well as higher rank gauge theories \cite{Vijay2016-dr,Slagle2017-ne,Ma2017-qq,Halasz2017-ov,Hsieh2017-sc,Vijay2017-ey,Slagle2017-gk,Ma2017-cb,Chamon2005-fc,yoshida2013exotic,Haah2011-ny,Slagle2017-la,shirley2017fracton,pretko2017fracton,ma2018fracton,prem2017emergent,pretko2017subdimensional,Ma2017-qq,bulmash2018higgs,Prem2017-ql,2018arXiv180601855B,you2018subsystem,devakul2018fractal,you2018symmetric}. Earlier literature shows that fracton order shares many features of topological order, including long-range entangled ground states and non-trivial braiding statistics. At the same time, fracton phases have a subextensive ground-state degeneracy depending on system size and lattice topology which transcends the paradigm of topological quantum field theory. Its quasiparticles have restricted mobility and move within lower-dimensional manifolds only. 

Interacting HOTSC may provide a new route towards realizing stabilizer codes of quantum information theory, with possible applications to topological quantum computation. In particular, the relation between interacting HOTSC and fracton phases may offer an inroad towards the experimental realization of fracton matter which remains little explored \cite{slagle2017fracton,Halasz2017-ov}. One possible platform uses Majorana wires where the interaction can be implemented as a charging energy \cite{landau2016towards}, allowing one to tune through and probe the quantum phase transition between HOTSC and topologically ordered phase. It has also been proposed that Sr$_2$RuO$_4$ \cite{raghu2010hidden} can be understood starting from a HOTSC model of the kind that we explore in this paper. 

\begin{figure}[t]
  \centering
      \includegraphics[width=0.25\textwidth]{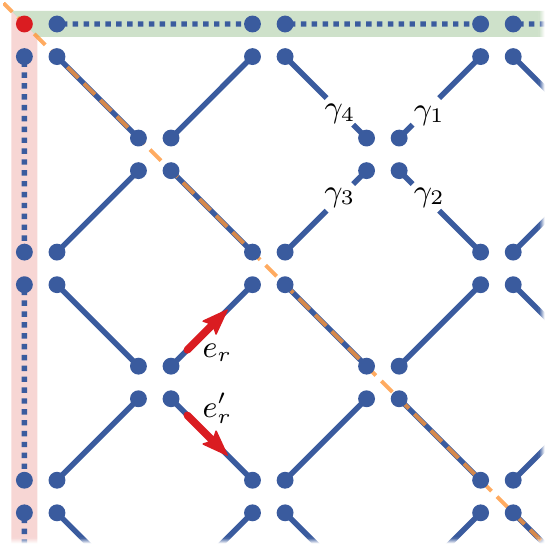}
  \caption{Model of HOTSC, see Eq.\ (\ref{H}). Each Majorana (blue dots) hybridizes with its nearest neighbor as indicated by the blue solid lines. The existence of Majorana corner modes (red dot) is protected by reflection symmetry about the dashed orange line. The products over the Majoranas enclosed by the green and red boxes are the fermion parities $P_x$ and $P_y$ of the edges.} 
  \label{levin1}
\end{figure}

{\em Model of HOTSC.---}We begin with a non-interacting band model on a square lattice with four Majoranas per site. Each Majorana hybridizes with its closest neighbor, as shown in Fig.~\ref{levin1} and described by the Hamiltonian 
\begin{align} 
H=-it \sum_j  (\gamma_j^1 \gamma_{j+e_r}^3+ \gamma_j^2 \gamma_{j+e_r'}^4).
\label{H}
\end{align}
Here, $j$ enumerates the lattice sites connected by lattice vectors $e_r$ and $e_r'$. Interestingly, this quasi-1D pairing structure has been proposed to describe superconductivity in Sr$_{2}$RuO$_4$, where it emerges from the directionality of $d_{xz}$ and $d_{yz}$ orbitals \cite{raghu2010hidden,benalcazar2014classification}.

With terminations along the $x$ and $y$ boundaries, all edge sites contain two unpaired Majorana zero modes (MZM). The edges can be gapped by hybridizing the MZM on the boundary with their nearest neighbor. This leaves one MZM uncoupled at the corner, see Fig.~\ref{levin1}. When imposing reflection symmetry about the diagonal axis passing through the corner or a $C_4$ rotation symmetry,
the MZM at the corner is symmetry protected, and the model describes a HOTSC. 

To demonstrate the robustness of the MZM at arbitrary reflection-symmetric corners, we define the operators $P_x=\prod_{j \in {\rm green}} \gamma_j$ and $P_y=\prod_{j \in {\rm red}} \gamma_j$ which measure the fermion parities of the dangling Majoranas on the $x$ and $y$ edges, as shown in Fig.~\ref{levin1}. Clearly, the operators $P_x$ and $P_y$ are connected by reflection $R$ and anticommute. Suppose now that a patch of HOTSC had a unique ground state $|\psi\rangle$. It must be reflection invariant and an eigenstate of $P_x P_y$ with eigenvalue $c$. The obstruction 
\begin{align} 
c| \psi \rangle = R  P_x  P_y | \psi \rangle =  P_y  P_x | \psi \rangle = -  P_x  P_y | \psi \rangle = - c| \psi \rangle
\label{parity}
\end{align}
implies that there is no unique reflection-symmetric ground state and the corners must carry protected zero modes. If we perturb away from the zero-correlation length Hamiltonian in Eq.~(\ref{H}), the argument survives provided that $P_x$ and $P_y$ are replaced by generalized edge parity operators, see supplementary material (SM) \cite{SM}.

\begin{figure}[t]
  \centering
      \includegraphics[width=0.36\textwidth]{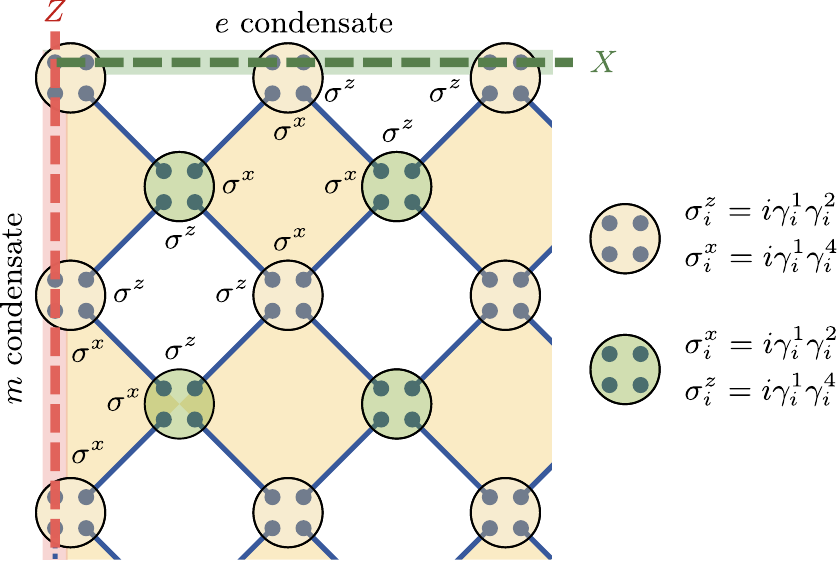}
  \caption{Surface code as a low-energy theory for the interacting HOTSC. Each site supports a spin 1/2 and the plaquette operators involve product of $\sigma^z$ ($\sigma^x$) for white (yellow) plaquettes. Three-spin interactions along the $x$ ($y$) edges induce $e$ ($m$)-particle condensation on the boundaries. The anti-commutation relation between the two Wilson lines (dashed red and green) implies a twofold topological ground-state degeneracy.}
  \label{levin2}
\end{figure}

{\em Interactions and topological order.---}We now turn on the onsite interaction 
\begin{equation}
   H_{\rm int} = U \sum_j \gamma^1_j\gamma^2_j\gamma^3_j\gamma^4_j.
\end{equation}
For each site, large $U$ fixes the fermion parity to $\gamma^1_j\gamma^2_j\gamma^3_j\gamma^4_j=-1$ and reduces the low-energy Hilbert space to a two-level system. The Pauli operators of this effective spin-${1}/{2}$ degree of freedom can be defined as shown in Fig.~\ref{levin2}. Performing a perturbation expansion in the Majorana hybridizations which flip the fermion parities of the participating sites, the leading-order effective Hamiltonian involves Majorana hopping terms around all elementary plaquettes. In terms of spin operators, this is just the $Z_2$ surface code with plaquette terms forming a checkerboard pattern and becoming $\prod_{\square} \sigma^z_i \sigma^z_j \sigma^z_k \sigma^z_l$ and $\prod_{\square}\sigma^x_i \sigma^x_j \sigma^x_k \sigma^x_l$ for white and yellow plaquettes, respectively (see Fig.~\ref{levin2}). With periodic boundary conditions, the resulting quantum stabilizer code is topologically ordered and its topological ground state degeneracy reflects global flux configurations. With open boundary conditions, the ground state can be unique for appropriate edge stabilizers, e.g., when condensing $e$ (or $m$) anyons along the entire boundary.

However, symmetries can impose a protected ground state degeneracy even for open patches. This follows directly from the fact that the mean-field Hamiltonian is a HOTSC. Its reflection symmetry interchanges not only lattice sites, but also the Pauli $X$ and $Z$ operators, $\sigma^x(x,y)\leftrightarrow \sigma^z(y,x)$, so that reflection maps $e$ into $m$-anyons and vice versa. Thus, $e$-anyon condensates on the top and bottom edges must come with $m$-anyon condensates on the left and right edges. Indeed, upon projection, the edge terminations of the HOTSC generate additional $\prod_{\triangledown}\sigma^x_i \sigma^x_j \sigma^x_k$ terms for the $x$-edge and $\prod_{ \triangleright}\sigma^z_i \sigma^z_j \sigma^z_k$ terms for the $y$-edge. These prompt the $e$ ($m$)-anyon condensation associated with rough (smooth) edges so that the surface code patch has a twofold ground-state degeneracy. 

\begin{figure}[t]
  \centering
      \includegraphics[width=0.4\textwidth]{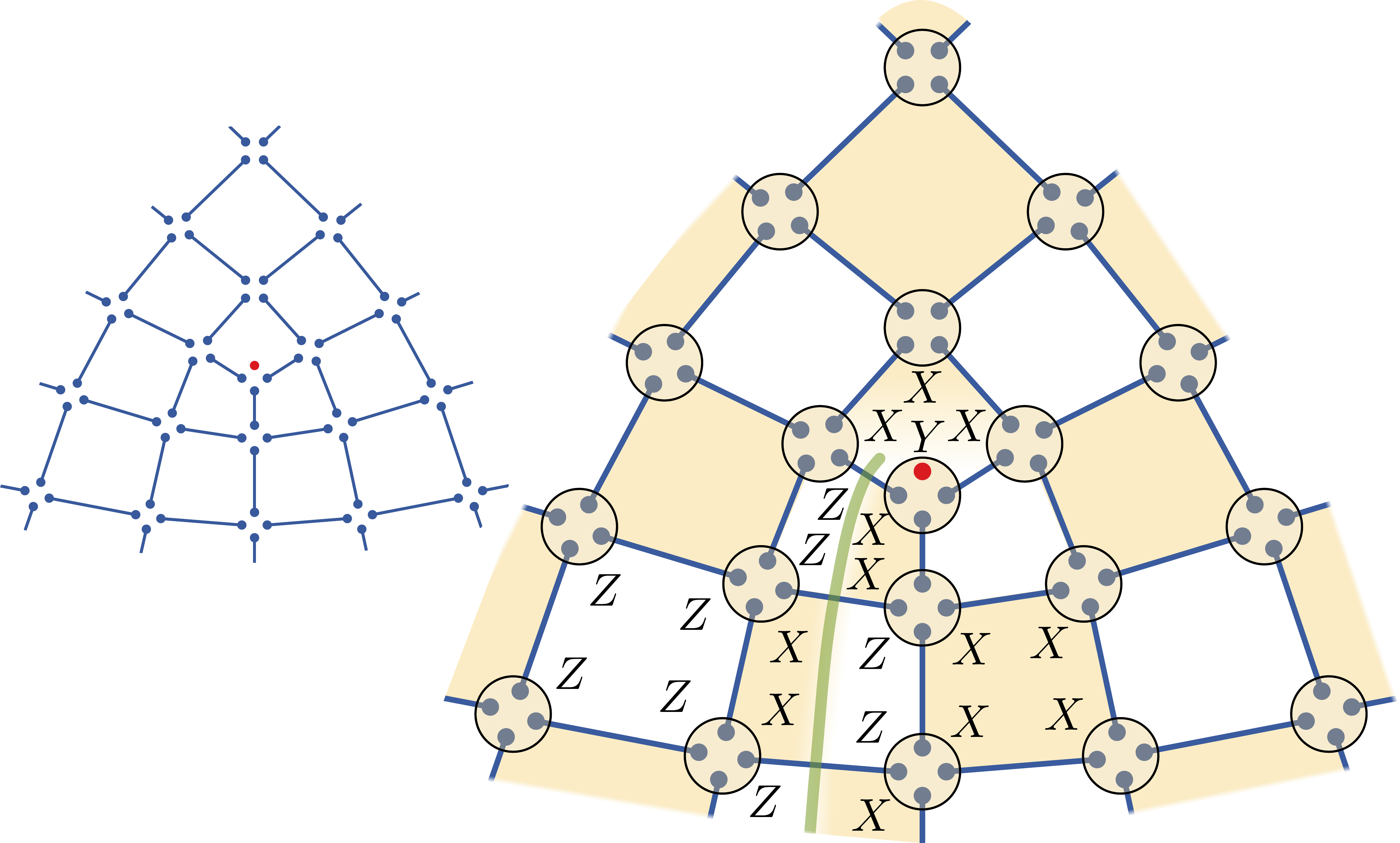}
  \caption{Left: A site-centered disclination of a HOTSC carrying a MZM. Right: In the strong interaction limit, the resulting toric code carries a $\epsilon$ fermion zero mode on the site-centered disclination. The operators on each plaquette specify the corresponding stabilizers.} 
  \label{dis1}
\end{figure}

Corresponding logical Pauli operators can be chosen as Wilson line operators $W_x = \prod_{ x-{\rm edge}} \sigma^x$ and $W_y = \prod_{y-{\rm edge}}\sigma^z$. Just like the boundary parity operators $P_x$ and $P_y$, these Wilson line operators are related by reflection and anticommute. Indeed, the boundary parities $P_x$ and $P_y$ of the HOTSC project exactly into the Wilson line operators of the surface code patch (see SM for details \cite{SM}). 

Corners of surface code patches with junctions of rough and smooth edges condense both $e$ and $m$-anyons. 
Consequently, an $\epsilon$ fermion created in one corner can be moved to the opposite corner without cost in energy, implying the existence of a fermion zero mode at corners \footnote{Unlike the corner mode in HOTSC, this $\epsilon$ fermion zero mode does not carry fermion parity.}. Thus, this strongly interacting HOTSC realizes a $Z_2$ topologically ordered phase enriched by reflection or rotation symmetry.

The zero mode is intimately related to Ising anyons associated with twist defects of surface codes \cite{bombin2010topological,brown2017},  
although corners do not possess braiding properties. For boundary modes protected by a $C_4$ rotation, one can gauge this symmetry by introducing ${\pi}/{2}$ disclinations. Removing one quarter of a surface code patch and distorting the resulting lattice to reconnect the two cuts trades a corner for a site-centered disclination in the bulk, see Fig.~\ref{dis1}. For the mean-field HOTSC, such a disclination comes with a local MZM \cite{thorngren2018gauging,else2018}. 
After parity projection, the disclination core contains an $\epsilon$ zero mode which provides a twist between $e$ and $m$ strings  \cite{you2012projective,barkeshli2014symmetry}
and corresponds to an Ising anyon. Refs.~\cite{bombin2010topological,you2012projective,barkeshli2014symmetry} demonstrate that a dislocation in the toric code carries an Ising anyon. Dislocations with odd Burgers vector can be thought of as bound states of site- and plaquette-centered disclinations. As the plaquette-centered disclination is bicolorable, the twist and the Ising anyon are associated with the 
site-centered disclination. 

Other surface as well as color codes are known in the literature. We show in the SM \cite{SM} how several of these can also be obtained from HOTSC by introducing strong local interactions. 

\begin{figure}[t]
  \centering
      \includegraphics[width=0.42\textwidth]{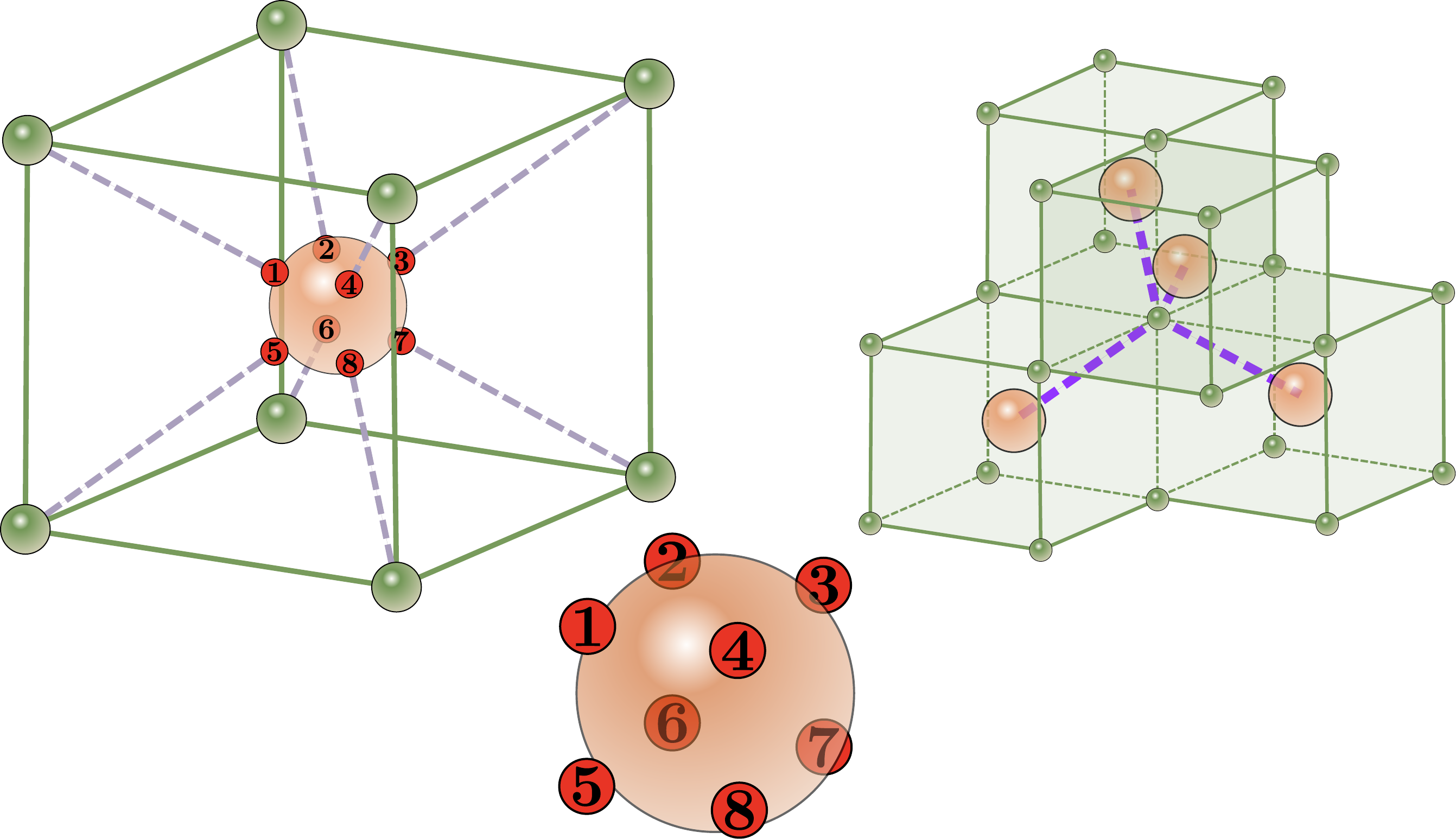}
  \caption{HOTSC on a checkerboard lattice. The vertices (green sites) host four Majoranas which are hybridized with the eight Majoranas at the center of the cubes (red sites). Majorana hybridization is indicated by dashed purple lines. After introducing local interactions, this HOTSC realizes a fracton code.} 
  \label{levin4}
\end{figure}

{\em Fracton codes.---}The emergence of quantum codes from interacting HOTSC can be extended to a variety of fracton codes \cite{unpub}.

We illustrate this connection for the specific case of a 3D HOTSC model on a body-centered checkerboard lattice, see Fig.~\ref{levin4}. Each vertex of the checkerboard lattice (green sites) hosts four Majoranas denoted by $\chi$. The center of the cubes (red sites) hosts eight Majoranas denoted by $\eta$. The Majoranas on the green sites pair with the closest Majorana on the neighboring red sites as shown in Fig.~\ref{levin4}. 
After projecting the hybridization structure into a plane,
the green sites on any 2d coordinate plane along with the red-site Majoranas coupled to them take just the form of the HOTSC on a square lattice shown in Fig.~\ref{levin1}.

In addition to the reflection symmetries, the Hamiltonian preserves a subsystem coplanar fermion parity for each layer, which counts the total fermion parity for the $\chi$ Majoranas of the layer together with the adjacent $\eta$ Majoranas coupled to them. The Hamiltonian can thus be considered as a 3D HOTSC protected by a subsystem symmetry \cite{you2018subsystem,devakul2018fractal,you2018symmetric,devakul2018strong}. Due to the subsystem symmetry, each plane contributes MZM at its corners, resulting in a line of MZM along the hinges of a finite cube. When maintaining both the subsystem fermion parity and the reflection symmetries for the $xy$ planes in a finite system with open (periodic) boundary conditions in the $x$ and $y$ ($z$) directions, we find a chain of protected MZM along the four $z$ hinges and hence a $2^{2L_z}$-fold degenerate ground state (see SM for details \cite{SM}).

\begin{figure}[t]
  \centering
      \includegraphics[width=0.4\textwidth]{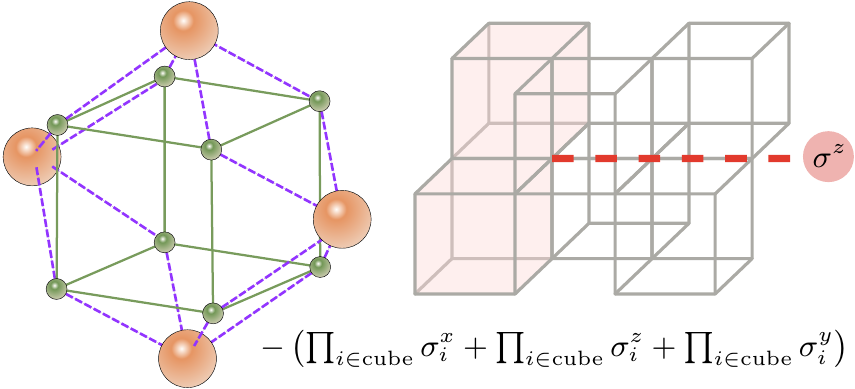}
  \caption{Checkerboard code. Left: In the strong interaction limit, the effective Hamiltonian involves products of 16 Majorana pairs (purple dashed lines) surrounding the four side faces of the cube. Each term can be expressed as an eight-spin cluster interaction on the checkerboard cube. Right: The spin Hamiltonian displays fractonic excitations where a pair of cube-flip excitations (red) can move only along a straight line by applying a $\sigma_z$ string operator.} 
    \label{levin5}
\end{figure}

We now introduce the onsite interaction
\begin{align} 
&H_{\eta}=U(\eta_1 \eta_2 \eta_3 \eta_4+\eta_5 \eta_6 \eta_7 \eta_8
+\eta_1 \eta_4 \eta_8 \eta_5\nonumber\\
&\,\,\,\,\,\,\,\,\,\,\,\,\,\,+\eta_2 \eta_3 \eta_7 \eta_6+\eta_2 \eta_1 \eta_5 \eta_6+\eta_3 \eta_4 \eta_8 \eta_7)
\label{omg}
\end{align}
on each red site, corresponding to four-Majorana interactions on each face of the cube, as well as a four-fermion interaction $H_{\chi}=U\chi_1 \chi_2 \chi_3\chi_4$ fixing the fermion parity for every green site. For strong interactions, the latter projects the four Majoranas of a green site into a spin-${1}/{2}$ degree of freedom which we denote by $\sigma$. The interaction $H_{\eta}$ projects the eight Majoranas on a red site into a unique state. As a result, the effective Hamiltonian in the strong interaction limit becomes
(see Fig.~\ref{levin5})
\begin{align} 
&H = - \sum_{\rm cubes} \left\{\prod_{j \in {\rm cube}}\sigma_j^x+\prod_{j \in {\rm cube}}\sigma_j^z + \prod_{j \in {\rm cube}}\sigma_j^y\right\},
\label{che}
\end{align}
which is known as the checkerboard model in the fracton literature \cite{Vijay2015-jj,Vijay2016-dr,Vijay2017-ey}. The interactions involve products of eight spins on all checkerboard cubes without red sites at the center. The red sites surrounding the cubes have no physical degree of freedom and merely act as gluons mediating the interaction between the spins $\sigma$. The fundamental excitation of the checkerboard model, a single cube flip, is completely immobile, while pairs of cube flips on adjacent planes are restricted to move along a fixed direction. For a cube of linear dimension $L$ with periodic boundary conditions, the ground state degeneracy is equal to $2^{6L-6}$ and different ground state sectors cannot be deformed into each other via local operators. 

For strong coupling, the coplanar fermion parity operators $P_x$ and $P_y$ on distinct edges project into the straight Wilson line operators $\prod_i \sigma^i_x$ and  $\prod_i \sigma^i_z$ which create pairs of cube flips limited to move along the $x$ or $y$ directions. While coplanar fermion parity operators determine the protected hinge modes of the HOTSC, the corresponding Wilson lines prompt the nontrivial braiding statistics between subdimensional particles and generate the Wilson algebra underlying its ground state degeneracy. Similar to surface code patches, one can explicitly relate the ground-state degeneracies of HOTSC and checkerboard model, see SM for details \cite{SM}. The reflection symmetry acts as a twist which permutes the two types of 1d subdimensional particles generated by the Pauli $X$ or $Z$ operators. In this sense, the reflection invariant checkerboard model is a symmetry-enriched fracton phase.

{\em Discussion.---}We have demonstrated that interactions can drive certain HOTSC into long-range entangled states with symmetry enriched topological or fracton order. This connection can be traced to the parton description of spin liquids which maps a strongly interacting boson system to a slave fermion theory coupled to a dynamical gauge field. At the mean-field level, the slave fermions obey a noninteracting band theory. Strong interactions between the slave fermions are mediated by the emergent gauge field which imposes a constraint on the local Hilbert space. For the $2D$ stabilizer codes discussed here, the interaction can be viewed as generators of local $Z_2$ gauge transformations. The interaction shares a single Majorana with the adjacent Majorana hybridization terms of the HOTSC Hamiltonian, which act as gauge connections. Hence, the generators flip the sign of the hybridization terms and for strong interactions, the effective low-energy Hamiltonian becomes a deconfined $Z_2$ gauge theory. Likewise, in the $3D$ fracton codes, the interaction  gauges the subsystem $Z_2$ symmetry for each plane, resulting in a higher-rank $Z_2$ gauge theory. This underlies the emergence of plaquette and cube operators characteristic of quantum code Hamiltonians. 

A crucial ingredient of the discussed HOTSC Hamiltonian is that the hybridization between Majoranas is quasi-one-dimensional. It has been proposed \cite{raghu2010hidden,benalcazar2014classification} that superconductivity in Sr$_2$RuO$_4$ can be described starting with the model shown in Fig.~\ref{levin1}. This may thus be a promising material basis for the physics discussed in this paper, although it may be challenging to substantially vary the strength of the onsite interaction experimentally. An alternative experimental platform relies on arrays of Majorana wires. The HOTSC Hamiltonians can be implemented with crossed Majorana wires, similar to the proposals in Ref.\ \cite{landau2016towards,vijay2015majorana,karzig2017scalable,2018arXiv180804529W,2018arXiv180709291T,2018arXiv180603304S}. Each cross carries four Majoranas at the wires' endpoints. The charging energy of a crossed Majorana wire with an epitaxial superconductor enforces a parity constraint, realizing an onsite four-Majorana interaction. The latter is experimentally tunable when coupling each Majorana cross to a bulk superconductor through a gate-tunable Josephson junction. Alternatively, one could tune the tunnel couplings between neighboring Majoranas, keeping the onsite interactions fixed. This tunability allows for experimental access to the quantum phase transition between the HOTSC and the toric code phase. One could further tune the onsite Majorana hybridizations to create synthetic twist defects which give rise to projective Ising anyons in the surface code \cite{you2012projective,you2012synthetic,barkeshli2013classification,bombin2010topological}.

A similar approach might also allow one to implement the checkerboard fracton model. In the SM \cite{SM}, we discuss how the more involved onsite interaction on the red sites with eight Majoranas could be implemented experimentally. 

{\em Acknowledgments.---}We acknowledge helpful discussions with Luka Trifunovic. This work was supported in part by CRC 183 of the Deutsche Forschungsgemeinschaft (DL and FvO) and the PCTS fellowship (YY). One of us (FvO) is grateful for sabbatical support from the IQIM, an NSF physics frontier center funded in part by the Moore Foundation. YY and FvO performed part of this work at the Aspen Center for Physics, which is supported by National Science Foundation grant PHY-1607611.

\clearpage

\setcounter{figure}{0}
\setcounter{section}{0}
\setcounter{equation}{0}
\renewcommand{\theequation}{S\arabic{equation}}
\renewcommand{\thefigure}{S\arabic{figure}}

\onecolumngrid

\renewcommand{\Fig}[1]{\mbox{Fig.\unitspace\ref{Sfig:#1}}}

\renewcommand{\Figure}[1]{\mbox{Figure\unitspace\ref{Sfig:#1}}}

\newcommand{\vsigma}{\mbox{\boldmath $\sigma$}}

\section*{\Large{Supplemental Material}}

\vspace{0.7cm}

\section{HOTSC and surface/color codes}

We elaborate on several aspects of our discussion of surface codes and their connection to HOTSC in the main text. 

\subsection{Proof and extension of Eq.~(2)}

The argument in the main text assumes the zero correlation length limit and a specific hybridization of the boundary Majoranas. Here, we show how to generalize the argument. 

The operators $P_x=\prod_{j \in {\rm green}} \gamma_j$ and $P_y=\prod_{j \in {\rm red}} \gamma_j$ are related by reflection $R$ and anticommute. (Here, we take $R$ to map Majorana operators at positions $j$ and $Rj$ related by reflection onto one another, $R\gamma_jR^{-1} = \gamma_{Rj}$. In microscopic models of topological superconductors such as the Kitaev chain, a pure reflection may not be a good symmetry as it changes the sign of the $p$-wave pairing. In this case, the symmetry operator $R$ acting on Hilbert space should be understood as a concatenated operator involving both the reflection and a global gauge transformation.) For the specific edge termination of the zero correlation length model, $P_x P_y$ commutes with the Hamiltonian in Eq.~(1). 
When deforming the Hamiltonian beyond the zero-correlation length limit, this is no longer true. Nevertheless, a unique ground state would still be an eigenstate of a generalized operator $P^U_x P^U_y$ localized near the two orthogonal edges and the corresponding argument in Eq.~(2) remains valid.

Take $|\psi\rangle_0$ to be the ground state of the zero correlation-length model. When deforming the Hamiltonian away from this limit without closing the gap and without breaking the reflection symmetry, the resulting ground state $|\psi \rangle_1$ can be expressed as 
\begin{align} 
| \psi \rangle_1 = U | \psi \rangle_0
\end{align}
Here, $U$ is a reflection-symmetric local unitary transformation. As demonstrated in Ref.~\cite{chen2010local}, symmetric local unitary operators define the equivalence classes of symmetry protected topological (SPT) phases. Thus, any two ground states which belong to the same HOTSC phase are connected by such a symmetric local unitary transformation.

Acting with the edge parity operators on $| \psi \rangle_0$ and applying $U$, one finds
\begin{align} 
&U P_x P_y | \psi \rangle_0 =  U c | \psi \rangle_0 \nonumber\\
&\rightarrow  U P_x P_y U^{-1} U | \psi \rangle_0 = c | \psi \rangle_1 \nonumber\\
&\rightarrow U P_x P_y U^{-1}  | \psi \rangle_1 = c | \psi \rangle_1
\label{gparity}
\end{align}
Thus, we can define generalized edge parity operators $P^U_x$ and $P^U_y$ such that $P^U_x P^U_y=U P_x P_y U^{-1} $. As $U$ is local and reflection symmetric, the two generalized edge parity operators are still related by reflection and only involve fermions localized near the two orthogonal edges. We can then repeat the argument in Eq.~(2) with these generalized operators and our conclusion on a protected corner mode still applies.

It may also be useful to observe that by symmetry, the boundary Hamiltonian contains an odd number of Majoranas, so that there must be a MZM for any symmetry-preserving edge termination. Symmetry also implies that this MZM is localized near the corner.

\subsection{Surface code patch as a symmetry enriched topological phase}

\begin{figure}[h]
  \centering
      \includegraphics[width=0.6\textwidth]{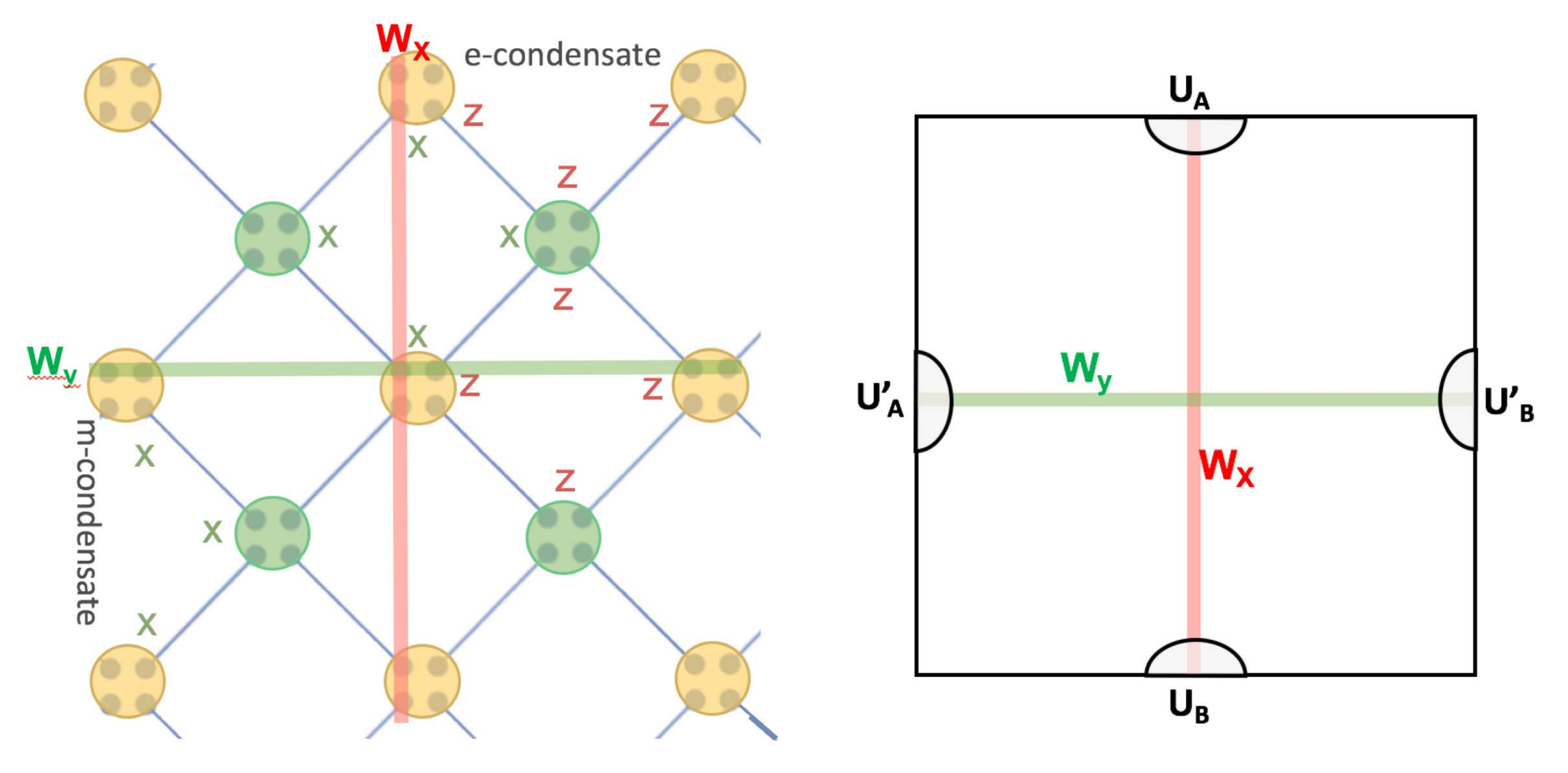} 
  \caption{Left: Reflection symmetric surface code with open boundary condition. The red ribbon ($e$ string) is mapped into the green ribbon ($m$ string) under reflection. Right: $e$/$m$ strings extending between boundaries, where the anyons can be annihilated via local operators.} 
    \label{fff}
\end{figure}

As discussed in the main text, the surface code can be regarded as a symmetry enriched topological phase where reflection acts as a twist defect permuting $e$ and $m$ anyons. With reflection symmetry, the surface code must have a ground state degeneracy for an open patch. The argument follows the discussion for the HOTSC and flows as follows. Quasiparticles of the $Z_2$ surface code contain a Lagrangian subgroup and the nontrivial quasiparticle in the Lagrangian subgroup can be annihilated at the gapped edge \cite{levin2013protected} via local operators. Without loss of generality, we take the $e$ anyon as the Lagrangian subgroup. We can then apply an $e$ string $W_x$ as shown in Fig.~\ref{fff} which ends on the top/bottom edge. Since $e$ anyons can be annihilated at the edge,  application of a local $U_A/U_B$ operator at the top/bottom edge returns the system to its ground state.

\begin{align} 
& U_A U_B W_x |\psi_{gs}\rangle=|\psi_{gs}\rangle.
\label{ls}
\end{align}
If the ground state was unique and reflection invariant, we could apply the same operation for horizontal $m$ strings as shown in Fig.~\ref{fff}. Applying the reflection operator on both sides of Eq.~(\ref{ls}) yields
\begin{align} 
& R~ U_A U_B W_x |\psi_{gs}\rangle=R |\psi_{gs}\rangle \rightarrow U'_A U'_B W_y |\psi_{gs}\rangle= |\psi_{gs}\rangle.
\end{align}
$ U'_A,U'_B$ are the reflection partners of $U_A,U_B$ which annihilate $m$ at the edge.
As $W_x W_y= -W_y W_x$, we find an obstruction,
\begin{align} 
& |\psi_{gs}\rangle = U_A U_B W_x U'_A U'_B W_y |\psi_{gs}\rangle= -U'_A U'_B W_y U_A U_B W_x |\psi_{gs}\rangle
=-|\psi_{gs}\rangle.
\end{align}
($ U'_A,U'_B$ and $U_A,U_B$ are local operators which always commute.) Hence, the reflection invariant toric code on an open patch must have a ground state degeneracy.

\subsection{Effective low energy Hamiltonian for the square lattice}

For completeness, we sketch the Brillouin-Wigner perturbation theory which yields the surface code from the strongly interacting HOTSC in Eq.~(1). Strong interactions enforce even fermion parity for each site. The effective Hamiltonian in this even-parity Hilbert space is obtained in third-order perturbation theory and takes the form
\begin{align}
H_{eff}=\sum_j -O(\frac{t^4}{U^3})\gamma^1_j \gamma^3_{j+e_r} \gamma^2_{j+e_r} \gamma^4_{j+e_r+e_{r'}} \gamma^2_j \gamma^4_{j+e_{r'}} \gamma^1_{j+e_{r'}} \gamma^3_{j+e_r+e_{r'}} 
\end{align}
Defining spin operators through $\sigma^z=i\gamma^1\gamma^2$ and $\sigma^x=i\gamma^1\gamma^4$ on yellow sites with the reverse definitions on green sites, the Hamiltonian becomes the surface code or Wen plaquette model on the checkerboard lattice,
\begin{align}
H_{eff}=-\sum_j O(\frac{t^4}{U^3}) (\prod_{j \in P_a} \sigma_j^z+\prod_{ j \in P_b}\sigma_j^x).
\end{align}

On the edge of the HOTSC, we consider nearest-neighbor tunneling between dangling Majoranas as shown Fig.~{1}. The effective Hamiltonian is now obtained from second-order perturbation, which yields
\begin{align}
H_{x-eff}=-\sum_j O(\frac{t^3}{U^2}) \prod_{ijk \in \Delta}\sigma_i^x \sigma_j^x\sigma_k^x
\end{align}
for the $x$-edge. This term prompts $e$-particle condensation on the $x$-edge. Similarly, the $y$-edge Hamiltonian becomes
\begin{align}
H_{y-eff}=-\sum_j O(\frac{t^3}{U^2}) \prod_{ijk \in \triangleright} \sigma_i^z \sigma_j^z \sigma_k^z,
\end{align}
which prompts $m$-particle condensation on the $y$-edge.

\subsection{Toric code and HOTSC on a Kagome lattice}

\begin{figure}[h]
  \centering
    \includegraphics[width=0.5\textwidth]{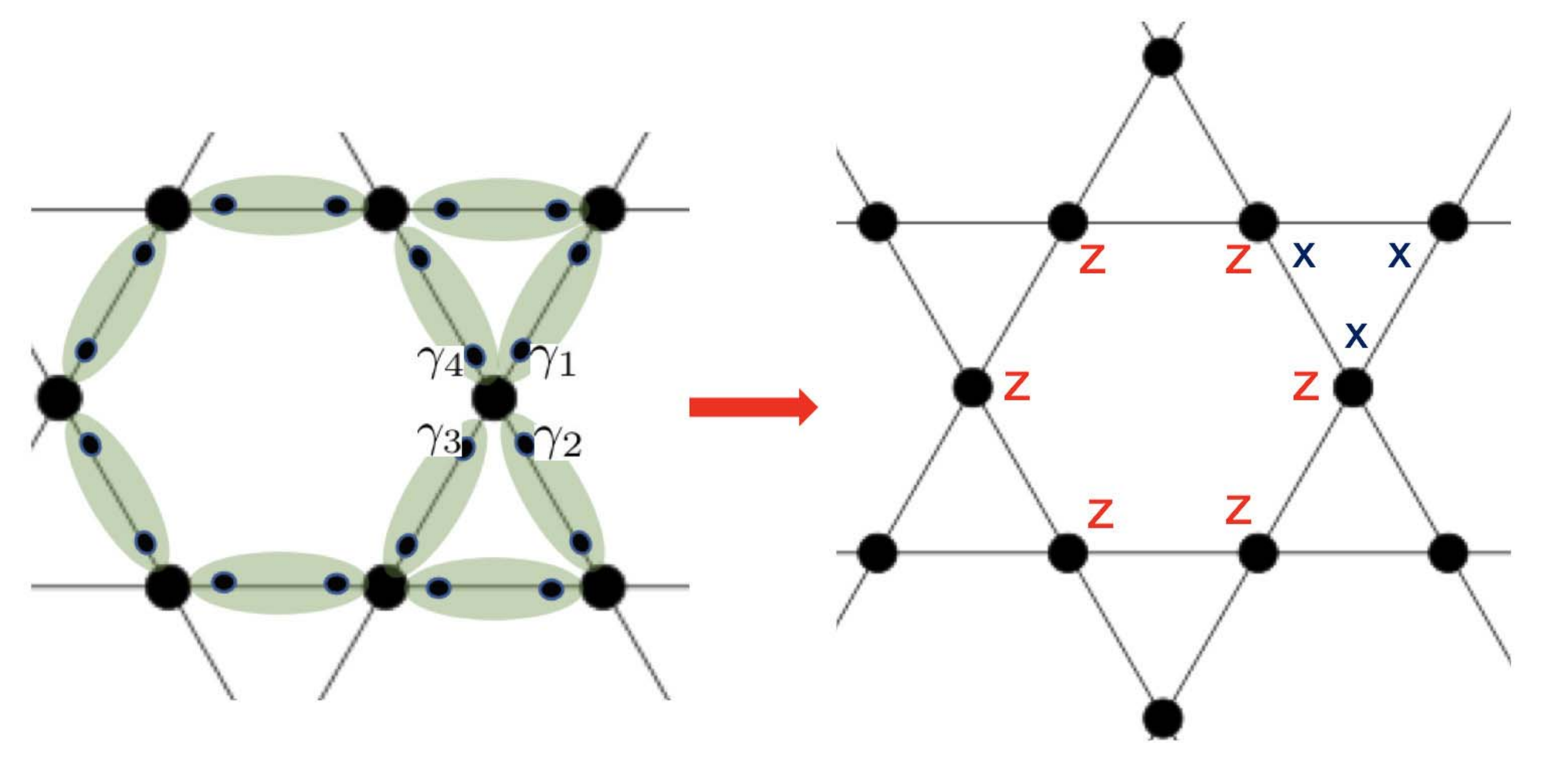} 
  \caption{HOTSC on a Kagome lattice with four Majorana per site. By onsite fermion parity fixing, the model is reduced to the toric code with spins located on the sites of a Kagome lattice (or equivalently, the bonds of a honeycomb lattice).} 
    \label{levinhere}
\end{figure}

In addition to the square lattice, we can also consider the surface code with qubits placed on the bonds of a honeycomb lattice, with plaquette operators involving six-spin terms and star operators corresponding to three-spin terms. The qubits are then located on the sites of a Kagome lattice. (This code is also equivalent to the 4.6.12 Majorana surface code.) This code can also be obtained from a HOTSC. Place crossing Kitaev wires along the links of a Kagome lattice with four Majoranas per site
as shown in Fig.~\ref{levinhere}. Onsite interactions $U \gamma^1 \gamma^2 \gamma^3 \gamma^4$ fix the local fermion parities, reducing each site to a spin-$1/2$ degree of freedom. The corresponding theory, obtained in second order and fifth order perturbation theory, is just the surface code on the Kagome lattice,
\begin{align}
H_{eff}=-\sum_j (O(\frac{t^6}{U^5}) \prod_{j \in \text{Hexagon}} \sigma_j^z+O(\frac{t^3}{U^2})\prod_{ j \in \text{Triangle}}\sigma_j^x)
\end{align}

\subsection{Majorana fermion code on honeycomb lattice}

\begin{figure}[h]
  \centering
      \includegraphics[width=0.45\textwidth]{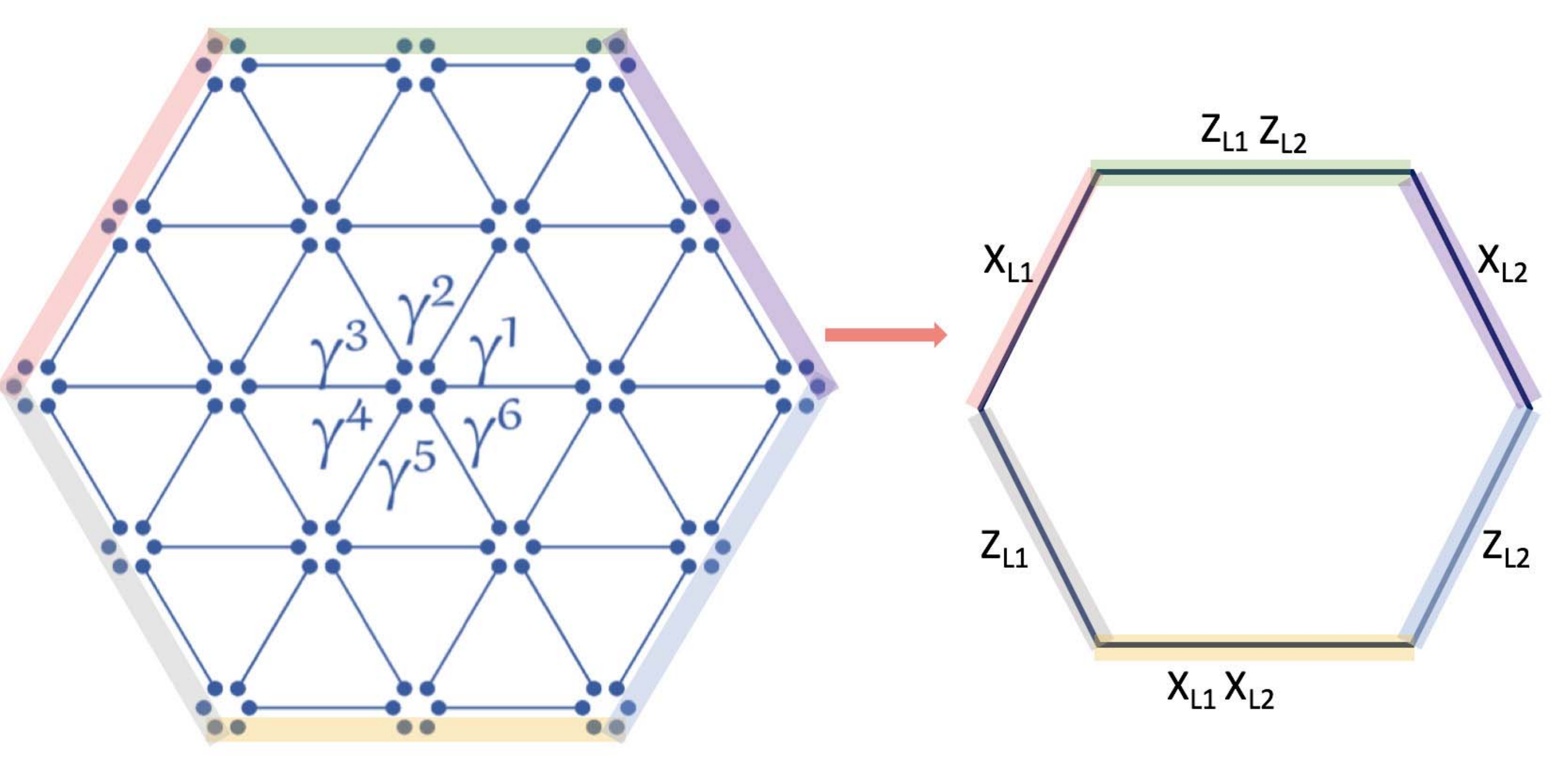}
  \caption{HOTSC and triangle code. Each site of the HOTSC hosts six Majoranas which hybridize with nearby sites as illustrated by the solid blue lines. Strong onsite interaction fixes the local fermion parity, turning the HOTSC into a triangle (or equivalently 6.6.6. Majorana fermion) code. Right: Possible definitions of two sets of Pauli $X$, $Y$, and $Z$ matrices associated with the fourfold degenerate ground state in the strong-interaction limit.} 
  \label{distantlight}
\end{figure}

As another example relating a HOTSC to a quantum code, consider a triangle lattice with six Majoranas on each site. The Majoranas hybridize with the nearest Majorana on a neighboring site as shown in Fig.~\ref{distantlight}. This HOTSC contains gapless Majorana corner modes protected by reflection symmetry about mirror axes placed at angles $\theta={2N\pi}/{3}$.

With strong onsite six-Majorana interactions 
 \begin{align} 
&H=- i U  \sum_j \eta_j^1 \eta_j^2 \eta_l^3 \eta_j^4 \eta_j^5 \eta_j^6,
\end{align}
the Hilbert space for each site is projected into a four-level system (representable by two Pauli spin operators) with fixed fermion parity.
Majorana hybridization is correspondingly suppressed and at low energies, the effective couplings are six-Majorana parity terms associated with each triangular lattice plaquette. Including the onsite interaction, the resulting model is just the 6.6.6 Majorana surface code on a honeycomb lattice \cite{vijay2015majorana,litinski2018quantum}. A hexagonal patch encodes two qubits with logical operators again emerging from the boundary parity operators of the HOTSC as indicated in Fig.\ref{distantlight}. 

\subsection{2D color codes}

HOTSC can also generate color codes 
\cite{bombin2006topological,landahl2011fault,Yoshida2013-of,vijay2015majorana}. Here, we show this explicitly for the color code on a honeycomb lattice.

Color codes are defined on three-colorable lattices so that plaquettes always share an even number of sites and both $X$ and $Z$ stabilizers can be defined on each plaquette. 
There are three types of $e$-particle excitations associated with plaquette flips on differently colored hexagons. These excitations are generated by three distinct string operators, with each string capable of branching into the other two.

\begin{figure}[h]
  \centering
    \includegraphics[width=0.27\textwidth]{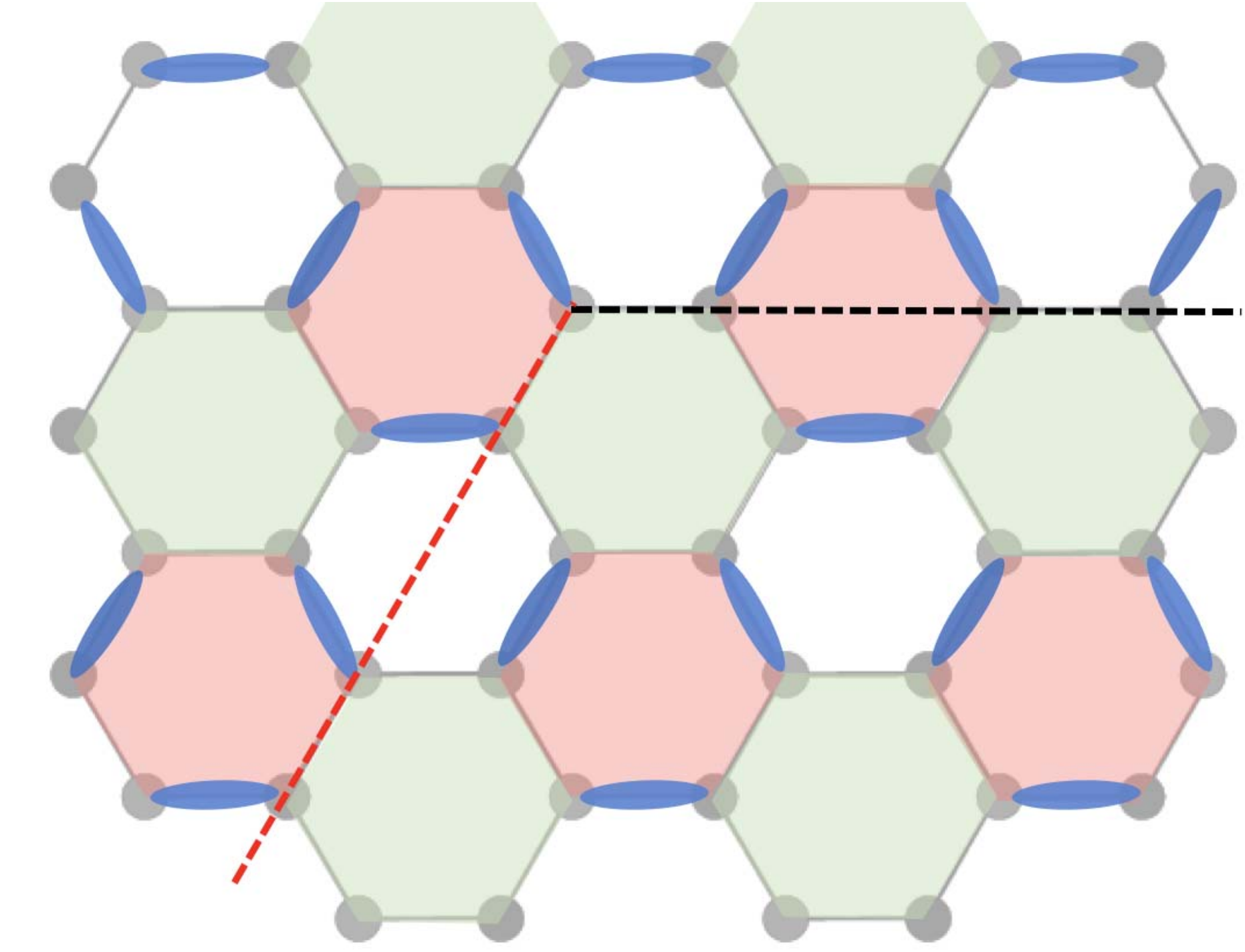} 
  \caption{HOTSC and color code. Each superconducting  island (green hexagons) has six MZM on its corner. Each Majorana hybridizes with its nearest neighbor as indicated by the blue bond.} 
    \label{color1}
\end{figure}
    
We initially follow the construction of the 6.6.6 Majorana fermion or triangle code described above, see Fig.~\ref{distantlight}. Majoranas on a honeycomb lattice hybridize as indicated by blue bonds in Fig.~\ref{color1}, forming three rotated Kitaev chains and obeying a discrete rotation symmetry. Interactions fix the fermion parity of the six Majoranas on the green hexagons. 
In the low-energy limit, the effective Hamiltonian is the 6.6.6 Majorana surface code whose stabilizers are products of Majoranas for every hexagon,
 \begin{align} 
  H=-i\sum_{\rm hexagons} \prod_{j \in \text{hex} } \eta_j.
\end{align}
Strictly speaking, stabilizers for green plaquettes (implemented as interactions) have a different amplitude from those on red and white plaquettes, which emerge due to Majorana hybridizations, but this does not affect the ground state subpace or the excitations. 

\begin{figure}[h]
  \centering
    \includegraphics[width=0.5\textwidth]{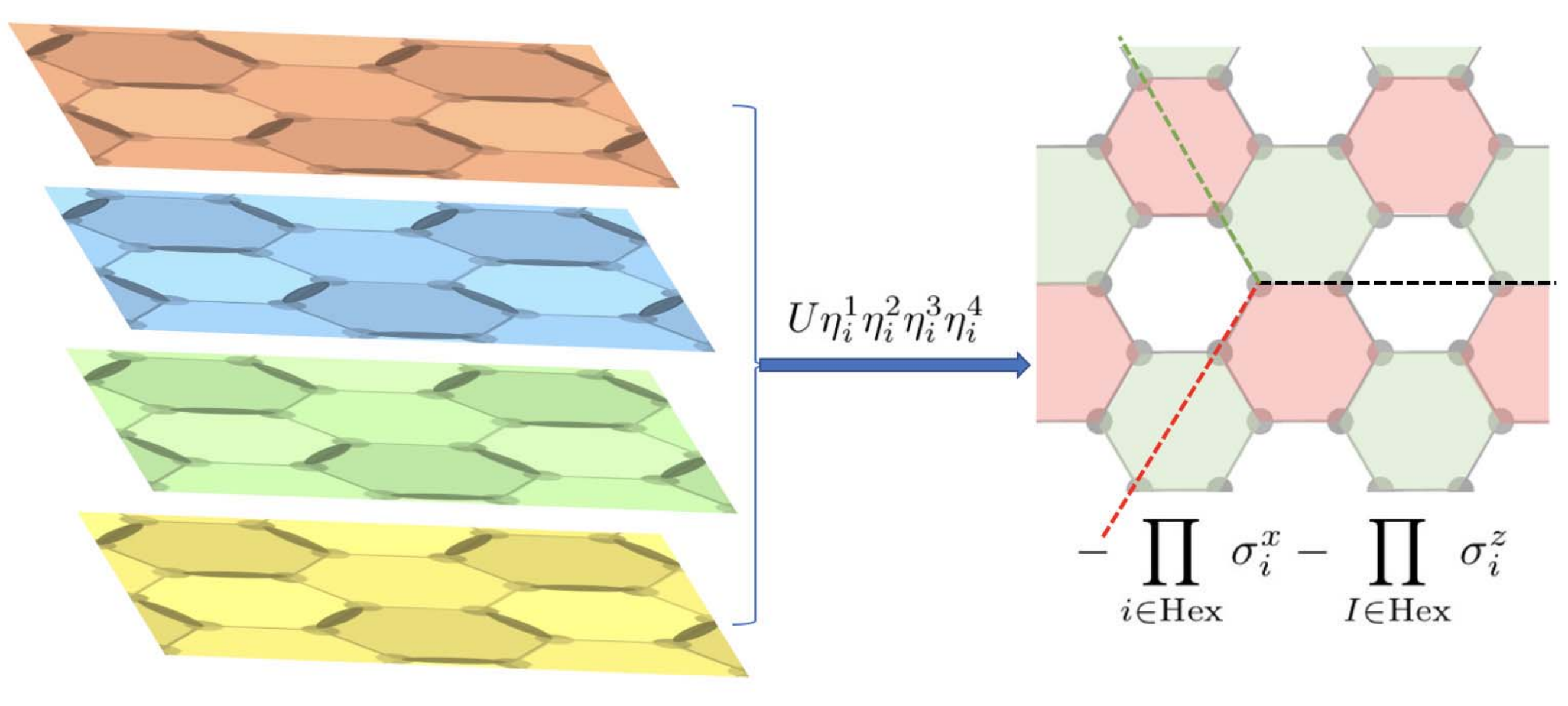} 
  \caption{Take four copies of the Majorana surface code and add interlayer interaction, the theory reduce to the color code on Honeycomb lattice. The three dashed lines are 3-types of string operators with one branching into the other two. } 
    \label{color2}
\end{figure}

We now take four copies of this construction as shown in Fig.~\ref{color2} and label the Majoranas of these layers by $\eta^1,\eta^2,\eta^3,\eta^4$. A strong onsite but interlayer interaction $U' \eta^1_j \eta^2_j  \eta^3_j  \eta^4_j$ reduces the low-energy Hilbert space on each site to a spin-$1/2$ degree of freedom and the resulting Hamiltonian becomes
 \begin{align} 
H=- \sum_{\rm hexagons} \left\{\prod_{i \in \text{hex} } \sigma^z_i+ \prod_{i \in \text{hex} } \sigma^x_i+ \prod_{i \in \text{hex} } \sigma^y_i\right\}
\label{Hcolor}
\end{align}
Indeed, with the interlayer interaction, the stabilizers of each layer no longer leave the system in its low-energy Hilbert space. This is only guaranteed for products of plaquette stabilizers which, when written in terms of Pauli operators for the site spins, just give the various terms in Eq.\ (\ref{Hcolor}). The last stabilizer in Eq.\ (\ref{Hcolor}) is redundant as it is simply the product of the first two. Thus, the Hamiltonian is exactly that of the color code on a honeycomb lattice. 
  
\section{HOTSC and fracton codes}

\subsection{Boundary parities and code space for checkerboard model}

\begin{figure}[h]
  \centering
      \includegraphics[width=0.36\textwidth]{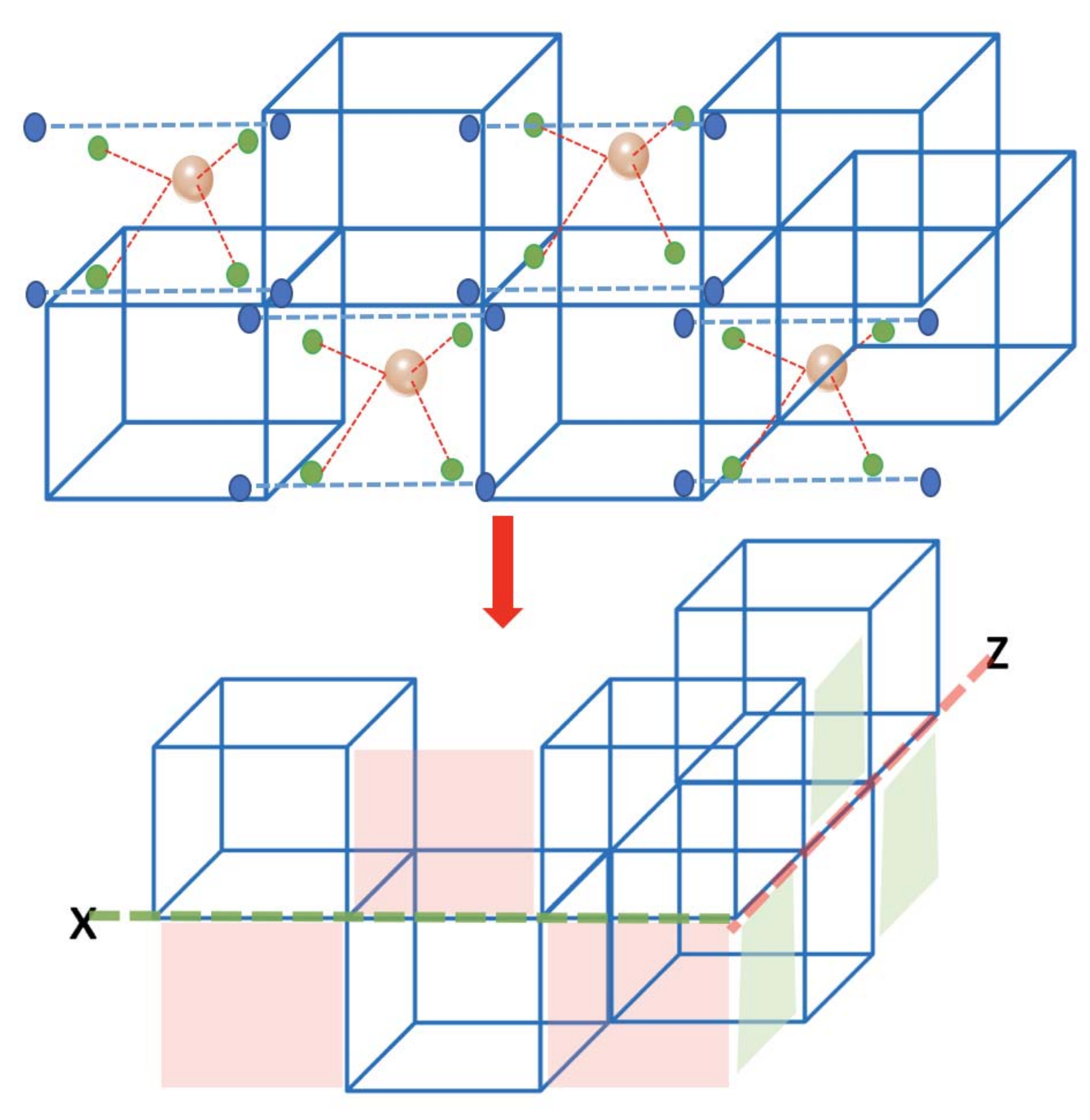}
  \caption{Top: The HOTSC with side faces on xz plane. The blue dots on the surfaces are the free MZM which can be gapped out via pairing(dashed blue lines). The yz plane can be gapped out in the similar way. Bottom: After projection, the surface Hamiltonian involves Z plaquette stabilizer(red) on xz side face and X plaquette stabilizer(red) on yz side face. The fermion parity operator becomes the Wilson straight line operators(dashed red and green line) along x or y directions.} 
  \label{relation}
\end{figure}

The argument connecting boundary parities and code space can be generalized to the checkerboard model. Start from the HOTSC in Fig.~{4} with open boundary conditions along the $x$ and $y$ directions and periodic boundary conditions along $z$. Then, the green sites on the side surfaces ($xz$ and $yz$ surfaces) contain two MZM which do not pair up. Similarly, each site on the $z$-hinge has three MZM. The subsystem fermion parity on each $xy$ plane along with the reflection symmetry (mapping between the $xz$ and $yz$ planes) protect one unpaired Majorana per $xy$ layer along each $z$ hinge, while the MZM on the side faces can be gapped out. Here, we specifically choose to gap out the surface MZM by pairing two nearby MZM along the $x$ ($y$) direction for the $xz$ ($yz$) plane as shown in Fig.~\ref{relation}. This surface termination does not break the subsystem fermion parity or the reflection symmetry of the $xy$ planes. As a result, there are extra dangling MZM at the hinges from each of the $L_z$ $xy$ layers, i.e., altogether $4L_z$ MZM resulting in a $2^{2L_z}$-fold degenerate ground state. As for the 2D HOTSC, we can define fermion parity operators
$P_x(z=z_i)$ and $P_y(z=z_i)$ which count the edge fermion parities for each $xy$ layer. 

Now we apply the projection induced by the onsite interactions [see Eq.~(4) and the accompanying discussion]. The resulting checkerboard model in Eq.~(5) now has additional boundary stabilizers
 \begin{align} 
& H_{xz-{\rm face}}=-\sum_{\rm red} \prod_{i \in {\rm red}} \sigma_i^z\,\,\,\, , \,\,\,\, H_{zy-{\rm face}}=-\sum_{\rm green}  \prod_{i \in {\rm green}} \sigma_i^x
\end{align}
On the $xz$ side face, the edge Hamiltonian reduces to $\sigma_z$ stabilizers on the red plaquettes. Similarly, it involves $\sigma_x$ stabilizers on the green plaquettes on the $yz$ side face. These surface stabilizers induce different (1d subdimensional) anyon condensates on the side surfaces. We can count the number of degrees of freedom and subtract the number of stabilizers of the Hamiltonian, yielding $L_z$. However, not all stabilizers are independent as the product of bulk and surface stabilizers along any $xz$ or $yz$ plane is equal to unity. Thus, we find a $2^{L_z+L_x+L_y-2}$-fold ground state degeneracy. (Note that we define $L_i$ as the number of layers of green sites along direction $i$.)  

On each $xy$ plane, the edge fermion parity operators $P_x(z=z_i)$ and $P_y(z=z_i)$ become the Wilson line operators $\prod_{i \in l_x}\sigma_i^x$ and $\prod_{i \in l_y}\sigma_i^z$ going along the $x$ and $y$ directions in the same plane. The commutation relations between these Wilson straight line operators indicate the braiding statistics of the $1d$ particles moving along the $x$ and $y$ directions. Due to the anyon condensation on $xz$ and $yz$ surfaces, the Wilson line operator $\prod_{i \in l_x}\sigma_i^x$ ($\prod_{i \in l_y}\sigma_i^z$) can extend across the entire $xz$ ($yz$) boundary without associated energy cost. This operation is a large gauge transformation, creating a global flux for each individual $xy$ plane and thus contributing a $2^{L_z}$-fold degeneracy. Now recall the fact that in the HOTSC, there are altogether $4L_z$ MZM along the four $z$-hinges. Fixing the fermion parity per $xy$ layer due to the onsite projection, the degeneracy associated with the hinges reduces to $2^{L_z}$ which exactly matches the result obtained by stabilizer counting or from the large gauge transformations of the fracton model. Finally, we note that the remaining $2^{L_x+L_y-2}$-fold degeneracy originates from the Wilson line algebra on the $xz$ or $yz$ planes.

Thus, the checkerboard code can be regarded as a symmetry enriched fracton phase. The reflection symmetry acts as a twist defect permuting $1d$ $e$ and $m$ particles about the mirror axis along the (110) direction. This reflection symmetry also permutes the $\sigma_x$  operator on the $xz$ surface and the $\sigma_z$ operator on the $yz$ surface. If we condense $e$ particles on $xz$ surfaces, there must be $m$ condensates on the $yz$ surfaces to satisfy the reflection symmetry constraint.

\subsection{Experimental implementation of interaction for fracton codes}

The transition between HOTSC and 2D surface code can be probed in arrays of Majorana wires. For instance, the HOTSC on a square lattice (see Fig.\ {1}) can be built from an array of crossed Majorana wires, with hybridization between nearest neighbor Majoranas on neighboring crosses. The four-Majorana parity constraint can be implemented through the charging energy of each Majorana cross with epitaxial superconductor. The charging energy is effectively tunable in experiment via the gate-tunable Josephson coupling of each Majorana cross to a bulk superconductor. Alternatively, one could tune the hybridization between neighboring Majoranas at fixed interaction strength. 

Our interacting HOTSC realizing the 3D checkerboard fracton model contains more complex interaction terms involving Majorana clusters with mutual overlap (red sites). Here we propose an experimental setup to realize this interaction. We choose the checkerboard code in Fig.~{4} for illustration, but the method applies more generally for a variety of fracton codes. 

\begin{figure}[h]
  \centering
    \includegraphics[width=0.2\textwidth]{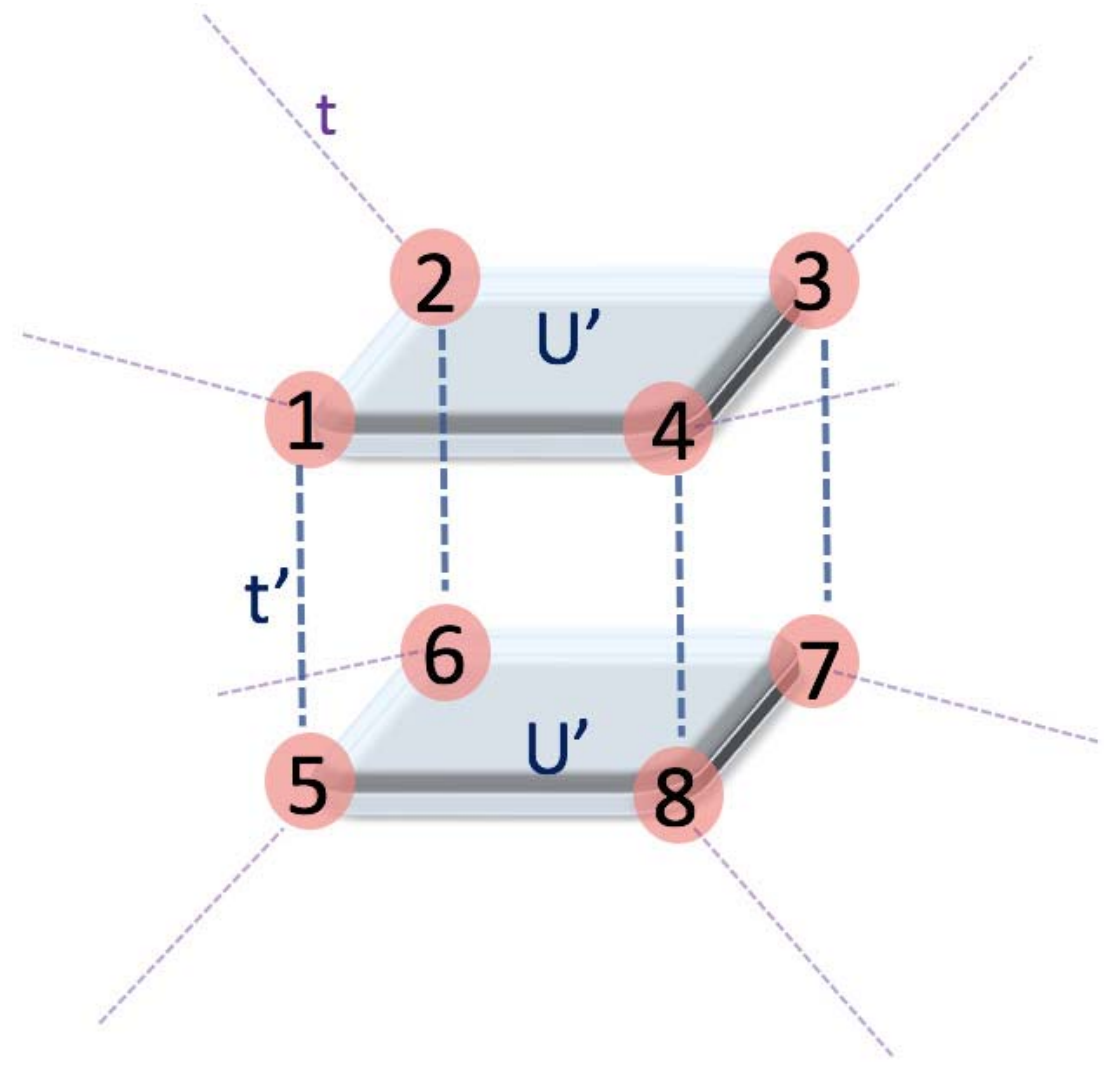} 
  \caption{There are two separate SC island(grey region) on each red site. The charging energy(U') fix the parity for $\eta_1 \eta_2 \eta_3 \eta_4$ and $\eta_5 \eta_6 \eta_7 \eta_8$ on each island. Tunneling between two island are denoted as the dashed blue lines. Such tunneling, in the strong U' limit, creates four Majorana interaction on the side faces.} 
    \label{exp}
\end{figure}

In the checkerboard model, the green sites contain four Majoranas and the corresponding interaction fixing the onsite parity can be implemented by placing the four Majoranas on a floating superconducting island. For the red sites at the center of the cubes, the interaction involves a four-fermion cluster interaction,
\begin{align} 
&H_{\eta}=U(\eta_1 \eta_2 \eta_3 \eta_4+\eta_5 \eta_6 \eta_7 \eta_8
+\eta_1 \eta_4 \eta_8 \eta_5+\eta_2 \eta_3 \eta_7 \eta_6+\eta_2 \eta_1 \eta_5 \eta_6+\eta_3 \eta_4 \eta_8 \eta_7)
\end{align}
There are only four independent interactions terms here and the rest can be obtained via product of the rest.
To engineer this interaction, we first place the Majorana $\eta_1, \eta_2, \eta_3 ,\eta_4$ and $\eta_5 ,\eta_6, \eta_7, \eta_8$ on two separate superconducting islands as shown in Fig.~\ref{exp}. The charging energies $U'(\eta_1 \eta_2 \eta_3 \eta_4+\eta_5 \eta_6 \eta_7 \eta_8)$ of these superconducting islands fix the parities $\eta_1 \eta_2 \eta_3 \eta_4$ and $\eta_5 \eta_6 \eta_7 \eta_8$. To generate the remaining four-Majorana interactions, we turn on inter-island hybridizations 
\begin{align} 
&H_{t'}=it'(\eta_1 \eta_5+\eta_2 \eta_6+\eta_3 \eta_7+\eta_4 \eta_8),
\end{align}
where $t'$ controls the hybridization. In the low-energy limit with fixed parities for the two islands, one then obtains the effective Hamiltonian
\begin{align} 
&H_{\rm eff}=U'(\eta_1 \eta_2 \eta_3 \eta_4+\eta_5 \eta_6 \eta_7 \eta_8)+
O(\frac{t'^2}{U'})(\eta_1 \eta_4 \eta_8 \eta_5+\eta_2 \eta_3 \eta_7 \eta_6+\eta_2 \eta_1 \eta_5 \eta_6+\eta_3 \eta_4 \eta_8 \eta_7 + \eta_1 \eta_3 \eta_5 \eta_7+\eta_2 \eta_4 \eta_8 \eta_6).
\end{align}
Except for the last two terms, these are just the interactions required on the red sites. Importantly, the anisotropy of the coefficients does not affect the ground state manifold. Similarly, the last two terms are products of two of the other terms and thus also do not affect the ground state manifold. When finally considering the effects of the hybridization $t$ between Majoranas on green and red sites to realize the fracton model, one has to ensure that one works in the regime $U'\gg t'\gg t$. This construction shows that a variety of bosonic fracton models, in the infrared limit, can be related to 3d Majorana codes. 


\begin{thebibliography}{79}%
\makeatletter
\providecommand \@ifxundefined [1]{%
 \@ifx{#1\undefined}
}%
\providecommand \@ifnum [1]{%
 \ifnum #1\expandafter \@firstoftwo
 \else \expandafter \@secondoftwo
 \fi
}%
\providecommand \@ifx [1]{%
 \ifx #1\expandafter \@firstoftwo
 \else \expandafter \@secondoftwo
 \fi
}%
\providecommand \natexlab [1]{#1}%
\providecommand \enquote  [1]{``#1''}%
\providecommand \bibnamefont  [1]{#1}%
\providecommand \bibfnamefont [1]{#1}%
\providecommand \citenamefont [1]{#1}%
\providecommand \href@noop [0]{\@secondoftwo}%
\providecommand \href [0]{\begingroup \@sanitize@url \@href}%
\providecommand \@href[1]{\@@startlink{#1}\@@href}%
\providecommand \@@href[1]{\endgroup#1\@@endlink}%
\providecommand \@sanitize@url [0]{\catcode `\\12\catcode `\$12\catcode
  `\&12\catcode `\#12\catcode `\^12\catcode `\_12\catcode `\%12\relax}%
\providecommand \@@startlink[1]{}%
\providecommand \@@endlink[0]{}%
\providecommand \url  [0]{\begingroup\@sanitize@url \@url }%
\providecommand \@url [1]{\endgroup\@href {#1}{\urlprefix }}%
\providecommand \urlprefix  [0]{URL }%
\providecommand \Eprint [0]{\href }%
\providecommand \doibase [0]{http://dx.doi.org/}%
\providecommand \selectlanguage [0]{\@gobble}%
\providecommand \bibinfo  [0]{\@secondoftwo}%
\providecommand \bibfield  [0]{\@secondoftwo}%
\providecommand \translation [1]{[#1]}%
\providecommand \BibitemOpen [0]{}%
\providecommand \bibitemStop [0]{}%
\providecommand \bibitemNoStop [0]{.\EOS\space}%
\providecommand \EOS [0]{\spacefactor3000\relax}%
\providecommand \BibitemShut  [1]{\csname bibitem#1\endcsname}%
\let\auto@bib@innerbib\@empty
\bibitem [{\citenamefont {Fu}(2011)}]{fu2011topological}%
  \BibitemOpen
  \bibfield  {author} {\bibinfo {author} {\bibfnamefont {L.}~\bibnamefont
  {Fu}},\ }\href@noop {} {\bibfield  {journal} {\bibinfo  {journal} {\prl}\
  }\textbf {\bibinfo {volume} {106}},\ \bibinfo {pages} {106802} (\bibinfo
  {year} {2011})}\BibitemShut {NoStop}%
\bibitem [{\citenamefont {Hsieh}\ \emph {et~al.}(2012)\citenamefont {Hsieh},
  \citenamefont {Lin}, \citenamefont {Liu}, \citenamefont {Duan}, \citenamefont
  {Bansil},\ and\ \citenamefont {Fu}}]{hsieh2012topological}%
  \BibitemOpen
  \bibfield  {author} {\bibinfo {author} {\bibfnamefont {T.~H.}\ \bibnamefont
  {Hsieh}}, \bibinfo {author} {\bibfnamefont {H.}~\bibnamefont {Lin}}, \bibinfo
  {author} {\bibfnamefont {J.}~\bibnamefont {Liu}}, \bibinfo {author}
  {\bibfnamefont {W.}~\bibnamefont {Duan}}, \bibinfo {author} {\bibfnamefont
  {A.}~\bibnamefont {Bansil}}, \ and\ \bibinfo {author} {\bibfnamefont
  {L.}~\bibnamefont {Fu}},\ }\href@noop {} {\bibfield  {journal} {\bibinfo
  {journal} {Nature Comm.}\ }\textbf {\bibinfo {volume} {3}},\ \bibinfo {pages}
  {982} (\bibinfo {year} {2012})}\BibitemShut {NoStop}%
\bibitem [{\citenamefont {Cheng}\ \emph {et~al.}(2016)\citenamefont {Cheng},
  \citenamefont {Zaletel}, \citenamefont {Barkeshli}, \citenamefont
  {Vishwanath},\ and\ \citenamefont {Bonderson}}]{cheng2016translational}%
  \BibitemOpen
  \bibfield  {author} {\bibinfo {author} {\bibfnamefont {M.}~\bibnamefont
  {Cheng}}, \bibinfo {author} {\bibfnamefont {M.}~\bibnamefont {Zaletel}},
  \bibinfo {author} {\bibfnamefont {M.}~\bibnamefont {Barkeshli}}, \bibinfo
  {author} {\bibfnamefont {A.}~\bibnamefont {Vishwanath}}, \ and\ \bibinfo
  {author} {\bibfnamefont {P.}~\bibnamefont {Bonderson}},\ }\href@noop {}
  {\bibfield  {journal} {\bibinfo  {journal} {Phys.\ Rev.\ X}\ }\textbf
  {\bibinfo {volume} {6}},\ \bibinfo {pages} {041068} (\bibinfo {year}
  {2016})}\BibitemShut {NoStop}%
\bibitem [{\citenamefont {Ando}\ and\ \citenamefont
  {Fu}(2015)}]{ando2015topological}%
  \BibitemOpen
  \bibfield  {author} {\bibinfo {author} {\bibfnamefont {Y.}~\bibnamefont
  {Ando}}\ and\ \bibinfo {author} {\bibfnamefont {L.}~\bibnamefont {Fu}},\
  }\href@noop {} {\bibfield  {journal} {\bibinfo  {journal} {Annu. Rev.
  Condens. Matter Phys.}\ }\textbf {\bibinfo {volume} {6}},\ \bibinfo {pages}
  {361} (\bibinfo {year} {2015})}\BibitemShut {NoStop}%
\bibitem [{\citenamefont {Slager}\ \emph {et~al.}(2013)\citenamefont {Slager},
  \citenamefont {Mesaros}, \citenamefont {Juri{\v{c}}i{\'c}},\ and\
  \citenamefont {Zaanen}}]{slager2013space}%
  \BibitemOpen
  \bibfield  {author} {\bibinfo {author} {\bibfnamefont {R.-J.}\ \bibnamefont
  {Slager}}, \bibinfo {author} {\bibfnamefont {A.}~\bibnamefont {Mesaros}},
  \bibinfo {author} {\bibfnamefont {V.}~\bibnamefont {Juri{\v{c}}i{\'c}}}, \
  and\ \bibinfo {author} {\bibfnamefont {J.}~\bibnamefont {Zaanen}},\
  }\href@noop {} {\bibfield  {journal} {\bibinfo  {journal} {Nature Physics}\
  }\textbf {\bibinfo {volume} {9}},\ \bibinfo {pages} {98} (\bibinfo {year}
  {2013})}\BibitemShut {NoStop}%
\bibitem [{\citenamefont {Hong}\ and\ \citenamefont
  {Fu}(2017)}]{hong2017topological}%
  \BibitemOpen
  \bibfield  {author} {\bibinfo {author} {\bibfnamefont {S.}~\bibnamefont
  {Hong}}\ and\ \bibinfo {author} {\bibfnamefont {L.}~\bibnamefont {Fu}},\
  }\href@noop {} {\bibfield  {journal} {\bibinfo  {journal} {arXiv:1707.02594}\
  } (\bibinfo {year} {2017})}\BibitemShut {NoStop}%
\bibitem [{\citenamefont {Qi}\ and\ \citenamefont
  {Fu}(2015)}]{qi2015anomalous}%
  \BibitemOpen
  \bibfield  {author} {\bibinfo {author} {\bibfnamefont {Y.}~\bibnamefont
  {Qi}}\ and\ \bibinfo {author} {\bibfnamefont {L.}~\bibnamefont {Fu}},\
  }\href@noop {} {\bibfield  {journal} {\bibinfo  {journal} {\prl}\ }\textbf
  {\bibinfo {volume} {115}},\ \bibinfo {pages} {236801} (\bibinfo {year}
  {2015})}\BibitemShut {NoStop}%
\bibitem [{\citenamefont {Huang}\ \emph {et~al.}(2017)\citenamefont {Huang},
  \citenamefont {Song}, \citenamefont {Huang},\ and\ \citenamefont
  {Hermele}}]{huang2017building}%
  \BibitemOpen
  \bibfield  {author} {\bibinfo {author} {\bibfnamefont {S.-J.}\ \bibnamefont
  {Huang}}, \bibinfo {author} {\bibfnamefont {H.}~\bibnamefont {Song}},
  \bibinfo {author} {\bibfnamefont {Y.-P.}\ \bibnamefont {Huang}}, \ and\
  \bibinfo {author} {\bibfnamefont {M.}~\bibnamefont {Hermele}},\ }\href@noop
  {} {\bibfield  {journal} {\bibinfo  {journal} {\prb}\ }\textbf {\bibinfo
  {volume} {96}},\ \bibinfo {pages} {205106} (\bibinfo {year}
  {2017})}\BibitemShut {NoStop}%
\bibitem [{\citenamefont {Teo}\ and\ \citenamefont
  {Hughes}(2013)}]{teo2013existence}%
  \BibitemOpen
  \bibfield  {author} {\bibinfo {author} {\bibfnamefont {J.~C.}\ \bibnamefont
  {Teo}}\ and\ \bibinfo {author} {\bibfnamefont {T.~L.}\ \bibnamefont
  {Hughes}},\ }\href@noop {} {\bibfield  {journal} {\bibinfo  {journal} {\prl}\
  }\textbf {\bibinfo {volume} {111}},\ \bibinfo {pages} {047006} (\bibinfo
  {year} {2013})}\BibitemShut {NoStop}%
\bibitem [{\citenamefont {Song}\ \emph
  {et~al.}(2017{\natexlab{a}})\citenamefont {Song}, \citenamefont {Huang},
  \citenamefont {Fu},\ and\ \citenamefont {Hermele}}]{song2017topological}%
  \BibitemOpen
  \bibfield  {author} {\bibinfo {author} {\bibfnamefont {H.}~\bibnamefont
  {Song}}, \bibinfo {author} {\bibfnamefont {S.-J.}\ \bibnamefont {Huang}},
  \bibinfo {author} {\bibfnamefont {L.}~\bibnamefont {Fu}}, \ and\ \bibinfo
  {author} {\bibfnamefont {M.}~\bibnamefont {Hermele}},\ }\href@noop {}
  {\bibfield  {journal} {\bibinfo  {journal} {Phys.\ Rev.\ X}\ }\textbf
  {\bibinfo {volume} {7}},\ \bibinfo {pages} {011020} (\bibinfo {year}
  {2017}{\natexlab{a}})}\BibitemShut {NoStop}%
\bibitem [{\citenamefont {Watanabe}\ \emph {et~al.}(2017)\citenamefont
  {Watanabe}, \citenamefont {Po},\ and\ \citenamefont
  {Vishwanath}}]{watanabe2017structure}%
  \BibitemOpen
  \bibfield  {author} {\bibinfo {author} {\bibfnamefont {H.}~\bibnamefont
  {Watanabe}}, \bibinfo {author} {\bibfnamefont {H.~C.}\ \bibnamefont {Po}}, \
  and\ \bibinfo {author} {\bibfnamefont {A.}~\bibnamefont {Vishwanath}},\
  }\href@noop {} {\bibfield  {journal} {\bibinfo  {journal} {arXiv:1707.01903}\
  } (\bibinfo {year} {2017})}\BibitemShut {NoStop}%
\bibitem [{\citenamefont {Po}\ \emph {et~al.}(2017)\citenamefont {Po},
  \citenamefont {Vishwanath},\ and\ \citenamefont {Watanabe}}]{po2017symmetry}%
  \BibitemOpen
  \bibfield  {author} {\bibinfo {author} {\bibfnamefont {H.~C.}\ \bibnamefont
  {Po}}, \bibinfo {author} {\bibfnamefont {A.}~\bibnamefont {Vishwanath}}, \
  and\ \bibinfo {author} {\bibfnamefont {H.}~\bibnamefont {Watanabe}},\
  }\href@noop {} {\bibfield  {journal} {\bibinfo  {journal} {Nature Comm.}\
  }\textbf {\bibinfo {volume} {8}},\ \bibinfo {pages} {50} (\bibinfo {year}
  {2017})}\BibitemShut {NoStop}%
\bibitem [{\citenamefont {Isobe}\ and\ \citenamefont
  {Fu}(2015)}]{isobe2015theory}%
  \BibitemOpen
  \bibfield  {author} {\bibinfo {author} {\bibfnamefont {H.}~\bibnamefont
  {Isobe}}\ and\ \bibinfo {author} {\bibfnamefont {L.}~\bibnamefont {Fu}},\
  }\href@noop {} {\bibfield  {journal} {\bibinfo  {journal} {\prb}\ }\textbf
  {\bibinfo {volume} {92}},\ \bibinfo {pages} {081304} (\bibinfo {year}
  {2015})}\BibitemShut {NoStop}%
\bibitem [{\citenamefont {Benalcazar}\ \emph
  {et~al.}(2017{\natexlab{a}})\citenamefont {Benalcazar}, \citenamefont
  {Bernevig},\ and\ \citenamefont {Hughes}}]{benalcazar2017quantized}%
  \BibitemOpen
  \bibfield  {author} {\bibinfo {author} {\bibfnamefont {W.~A.}\ \bibnamefont
  {Benalcazar}}, \bibinfo {author} {\bibfnamefont {B.~A.}\ \bibnamefont
  {Bernevig}}, \ and\ \bibinfo {author} {\bibfnamefont {T.~L.}\ \bibnamefont
  {Hughes}},\ }\href@noop {} {\bibfield  {journal} {\bibinfo  {journal}
  {Science}\ }\textbf {\bibinfo {volume} {357}},\ \bibinfo {pages} {61}
  (\bibinfo {year} {2017}{\natexlab{a}})}\BibitemShut {NoStop}%
\bibitem [{\citenamefont {Benalcazar}\ \emph
  {et~al.}(2017{\natexlab{b}})\citenamefont {Benalcazar}, \citenamefont
  {Bernevig},\ and\ \citenamefont {Hughes}}]{benalcazar2017electric}%
  \BibitemOpen
  \bibfield  {author} {\bibinfo {author} {\bibfnamefont {W.~A.}\ \bibnamefont
  {Benalcazar}}, \bibinfo {author} {\bibfnamefont {B.~A.}\ \bibnamefont
  {Bernevig}}, \ and\ \bibinfo {author} {\bibfnamefont {T.~L.}\ \bibnamefont
  {Hughes}},\ }\href@noop {} {\bibfield  {journal} {\bibinfo  {journal} {\prb}\
  }\textbf {\bibinfo {volume} {96}},\ \bibinfo {pages} {245115} (\bibinfo
  {year} {2017}{\natexlab{b}})}\BibitemShut {NoStop}%
\bibitem [{\citenamefont {Langbehn}\ \emph {et~al.}(2017)\citenamefont
  {Langbehn}, \citenamefont {Peng}, \citenamefont {Trifunovic}, \citenamefont
  {von Oppen},\ and\ \citenamefont {Brouwer}}]{langbehn2017reflection}%
  \BibitemOpen
  \bibfield  {author} {\bibinfo {author} {\bibfnamefont {J.}~\bibnamefont
  {Langbehn}}, \bibinfo {author} {\bibfnamefont {Y.}~\bibnamefont {Peng}},
  \bibinfo {author} {\bibfnamefont {L.}~\bibnamefont {Trifunovic}}, \bibinfo
  {author} {\bibfnamefont {F.}~\bibnamefont {von Oppen}}, \ and\ \bibinfo
  {author} {\bibfnamefont {P.~W.}\ \bibnamefont {Brouwer}},\ }\href@noop {}
  {\bibfield  {journal} {\bibinfo  {journal} {\prl}\ }\textbf {\bibinfo
  {volume} {119}},\ \bibinfo {pages} {246401} (\bibinfo {year}
  {2017})}\BibitemShut {NoStop}%
\bibitem [{\citenamefont {Song}\ \emph
  {et~al.}(2017{\natexlab{b}})\citenamefont {Song}, \citenamefont {Fang},\ and\
  \citenamefont {Fang}}]{song2017d}%
  \BibitemOpen
  \bibfield  {author} {\bibinfo {author} {\bibfnamefont {Z.}~\bibnamefont
  {Song}}, \bibinfo {author} {\bibfnamefont {Z.}~\bibnamefont {Fang}}, \ and\
  \bibinfo {author} {\bibfnamefont {C.}~\bibnamefont {Fang}},\ }\href@noop {}
  {\bibfield  {journal} {\bibinfo  {journal} {\prl}\ }\textbf {\bibinfo
  {volume} {119}},\ \bibinfo {pages} {246402} (\bibinfo {year}
  {2017}{\natexlab{b}})}\BibitemShut {NoStop}%
\bibitem [{\citenamefont {Dwivedi}\ \emph {et~al.}(2018)\citenamefont
  {Dwivedi}, \citenamefont {Hickey}, \citenamefont {Eschmann},\ and\
  \citenamefont {Trebst}}]{dwivedi2018majorana}%
  \BibitemOpen
  \bibfield  {author} {\bibinfo {author} {\bibfnamefont {V.}~\bibnamefont
  {Dwivedi}}, \bibinfo {author} {\bibfnamefont {C.}~\bibnamefont {Hickey}},
  \bibinfo {author} {\bibfnamefont {T.}~\bibnamefont {Eschmann}}, \ and\
  \bibinfo {author} {\bibfnamefont {S.}~\bibnamefont {Trebst}},\ }\href@noop {}
  {\bibfield  {journal} {\bibinfo  {journal} {arXiv:1803.08922}\ } (\bibinfo
  {year} {2018})}\BibitemShut {NoStop}%
\bibitem [{\citenamefont {Schindler}\ \emph {et~al.}(2018)\citenamefont
  {Schindler}, \citenamefont {Cook}, \citenamefont {Vergniory}, \citenamefont
  {Wang}, \citenamefont {Parkin}, \citenamefont {Bernevig},\ and\ \citenamefont
  {Neupert}}]{schindler2018higher}%
  \BibitemOpen
  \bibfield  {author} {\bibinfo {author} {\bibfnamefont {F.}~\bibnamefont
  {Schindler}}, \bibinfo {author} {\bibfnamefont {A.~M.}\ \bibnamefont {Cook}},
  \bibinfo {author} {\bibfnamefont {M.~G.}\ \bibnamefont {Vergniory}}, \bibinfo
  {author} {\bibfnamefont {Z.}~\bibnamefont {Wang}}, \bibinfo {author}
  {\bibfnamefont {S.~S.}\ \bibnamefont {Parkin}}, \bibinfo {author}
  {\bibfnamefont {B.~A.}\ \bibnamefont {Bernevig}}, \ and\ \bibinfo {author}
  {\bibfnamefont {T.}~\bibnamefont {Neupert}},\ }\href@noop {} {\bibfield
  {journal} {\bibinfo  {journal} {Science Adv.}\ }\textbf {\bibinfo {volume}
  {4}},\ \bibinfo {pages} {eaat0346} (\bibinfo {year} {2018})}\BibitemShut
  {NoStop}%
\bibitem [{\citenamefont {Benalcazar}\ \emph {et~al.}(2014)\citenamefont
  {Benalcazar}, \citenamefont {Teo},\ and\ \citenamefont
  {Hughes}}]{benalcazar2014classification}%
  \BibitemOpen
  \bibfield  {author} {\bibinfo {author} {\bibfnamefont {W.~A.}\ \bibnamefont
  {Benalcazar}}, \bibinfo {author} {\bibfnamefont {J.~C.}\ \bibnamefont {Teo}},
  \ and\ \bibinfo {author} {\bibfnamefont {T.~L.}\ \bibnamefont {Hughes}},\
  }\href@noop {} {\bibfield  {journal} {\bibinfo  {journal} {\prb}\ }\textbf
  {\bibinfo {volume} {89}},\ \bibinfo {pages} {224503} (\bibinfo {year}
  {2014})}\BibitemShut {NoStop}%
\bibitem [{\citenamefont {Raghu}\ \emph {et~al.}(2010)\citenamefont {Raghu},
  \citenamefont {Kapitulnik},\ and\ \citenamefont
  {Kivelson}}]{raghu2010hidden}%
  \BibitemOpen
  \bibfield  {author} {\bibinfo {author} {\bibfnamefont {S.}~\bibnamefont
  {Raghu}}, \bibinfo {author} {\bibfnamefont {A.}~\bibnamefont {Kapitulnik}}, \
  and\ \bibinfo {author} {\bibfnamefont {S.}~\bibnamefont {Kivelson}},\
  }\href@noop {} {\bibfield  {journal} {\bibinfo  {journal} {\prl}\ }\textbf
  {\bibinfo {volume} {105}},\ \bibinfo {pages} {136401} (\bibinfo {year}
  {2010})}\BibitemShut {NoStop}%
\bibitem [{\citenamefont {Serra-Garcia}\ \emph {et~al.}(2018)\citenamefont
  {Serra-Garcia}, \citenamefont {Peri}, \citenamefont {S{\"u}sstrunk},
  \citenamefont {Bilal}, \citenamefont {Larsen}, \citenamefont {Villanueva},\
  and\ \citenamefont {Huber}}]{serra2018observation}%
  \BibitemOpen
  \bibfield  {author} {\bibinfo {author} {\bibfnamefont {M.}~\bibnamefont
  {Serra-Garcia}}, \bibinfo {author} {\bibfnamefont {V.}~\bibnamefont {Peri}},
  \bibinfo {author} {\bibfnamefont {R.}~\bibnamefont {S{\"u}sstrunk}}, \bibinfo
  {author} {\bibfnamefont {O.~R.}\ \bibnamefont {Bilal}}, \bibinfo {author}
  {\bibfnamefont {T.}~\bibnamefont {Larsen}}, \bibinfo {author} {\bibfnamefont
  {L.~G.}\ \bibnamefont {Villanueva}}, \ and\ \bibinfo {author} {\bibfnamefont
  {S.~D.}\ \bibnamefont {Huber}},\ }\href@noop {} {\bibfield  {journal}
  {\bibinfo  {journal} {Nature}\ }\textbf {\bibinfo {volume} {555}},\ \bibinfo
  {pages} {342} (\bibinfo {year} {2018})}\BibitemShut {NoStop}%
\bibitem [{\citenamefont {Bultinck}\ \emph {et~al.}()\citenamefont {Bultinck},
  \citenamefont {Bernevig},\ and\ \citenamefont {Zaletel}}]{andrei}%
  \BibitemOpen
  \bibfield  {author} {\bibinfo {author} {\bibfnamefont {N.}~\bibnamefont
  {Bultinck}}, \bibinfo {author} {\bibfnamefont {A.}~\bibnamefont {Bernevig}},
  \ and\ \bibinfo {author} {\bibfnamefont {M.}~\bibnamefont {Zaletel}},\
  }\href@noop {} {\bibinfo  {journal} {unpublished}\ }\BibitemShut {NoStop}%
\bibitem [{\citenamefont {Song}\ and\ \citenamefont
  {Schnyder}(2017)}]{song2017interaction}%
  \BibitemOpen
\bibfield  {journal} {  }\bibfield  {author} {\bibinfo {author} {\bibfnamefont
  {X.-Y.}\ \bibnamefont {Song}}\ and\ \bibinfo {author} {\bibfnamefont {A.~P.}\
  \bibnamefont {Schnyder}},\ }\href@noop {} {\bibfield  {journal} {\bibinfo
  {journal} {\prb}\ }\textbf {\bibinfo {volume} {95}},\ \bibinfo {pages}
  {195108} (\bibinfo {year} {2017})}\BibitemShut {NoStop}%
\bibitem [{\citenamefont {You}\ \emph {et~al.}(2018{\natexlab{a}})\citenamefont
  {You}, \citenamefont {Devakul}, \citenamefont {Burnell},\ and\ \citenamefont
  {Neupert}}]{you2018higher}%
  \BibitemOpen
  \bibfield  {author} {\bibinfo {author} {\bibfnamefont {Y.}~\bibnamefont
  {You}}, \bibinfo {author} {\bibfnamefont {T.}~\bibnamefont {Devakul}},
  \bibinfo {author} {\bibfnamefont {F.}~\bibnamefont {Burnell}}, \ and\
  \bibinfo {author} {\bibfnamefont {T.}~\bibnamefont {Neupert}},\ }\href@noop
  {} {\bibfield  {journal} {\bibinfo  {journal} {arXiv:1807.09788}\ } (\bibinfo
  {year} {2018}{\natexlab{a}})}\BibitemShut {NoStop}%
\bibitem [{\citenamefont {{Rasmussen}}\ and\ \citenamefont
  {{Lu}}(2018)}]{2018arXiv180907325R}%
  \BibitemOpen
  \bibfield  {author} {\bibinfo {author} {\bibfnamefont {A.}~\bibnamefont
  {{Rasmussen}}}\ and\ \bibinfo {author} {\bibfnamefont {Y.-M.}\ \bibnamefont
  {{Lu}}},\ }\href@noop {} {\bibfield  {journal} {\bibinfo  {journal}
  {arxiv:1809.07325}\ } (\bibinfo {year} {2018})}\BibitemShut {NoStop}%
\bibitem [{\citenamefont {Wen}(2003)}]{wen2003quantum}%
  \BibitemOpen
  \bibfield  {author} {\bibinfo {author} {\bibfnamefont {X.-G.}\ \bibnamefont
  {Wen}},\ }\href@noop {} {\bibfield  {journal} {\bibinfo  {journal} {\prl}\
  }\textbf {\bibinfo {volume} {90}},\ \bibinfo {pages} {016803} (\bibinfo
  {year} {2003})}\BibitemShut {NoStop}%
\bibitem [{\citenamefont {You}\ and\ \citenamefont
  {Wen}(2012)}]{you2012projective}%
  \BibitemOpen
  \bibfield  {author} {\bibinfo {author} {\bibfnamefont {Y.-Z.}\ \bibnamefont
  {You}}\ and\ \bibinfo {author} {\bibfnamefont {X.-G.}\ \bibnamefont {Wen}},\
  }\href@noop {} {\bibfield  {journal} {\bibinfo  {journal} {\prb}\ }\textbf
  {\bibinfo {volume} {86}},\ \bibinfo {pages} {161107} (\bibinfo {year}
  {2012})}\BibitemShut {NoStop}%
\bibitem [{\citenamefont {Hsieh}\ \emph {et~al.}(2017)\citenamefont {Hsieh},
  \citenamefont {Lu},\ and\ \citenamefont {Ludwig}}]{hsieh2017topological}%
  \BibitemOpen
  \bibfield  {author} {\bibinfo {author} {\bibfnamefont {T.~H.}\ \bibnamefont
  {Hsieh}}, \bibinfo {author} {\bibfnamefont {Y.-M.}\ \bibnamefont {Lu}}, \
  and\ \bibinfo {author} {\bibfnamefont {A.~W.}\ \bibnamefont {Ludwig}},\
  }\href@noop {} {\bibfield  {journal} {\bibinfo  {journal} {Science Adv.}\
  }\textbf {\bibinfo {volume} {3}},\ \bibinfo {pages} {e1700729} (\bibinfo
  {year} {2017})}\BibitemShut {NoStop}%
\bibitem [{\citenamefont {Vijay}\ \emph {et~al.}(2016)\citenamefont {Vijay},
  \citenamefont {Haah},\ and\ \citenamefont {Fu}}]{Vijay2016-dr}%
  \BibitemOpen
  \bibfield  {author} {\bibinfo {author} {\bibfnamefont {S.}~\bibnamefont
  {Vijay}}, \bibinfo {author} {\bibfnamefont {J.}~\bibnamefont {Haah}}, \ and\
  \bibinfo {author} {\bibfnamefont {L.}~\bibnamefont {Fu}},\ }\href@noop {}
  {\bibfield  {journal} {\bibinfo  {journal} {Phys. Rev. B}\ }\textbf {\bibinfo
  {volume} {94}},\ \bibinfo {pages} {235157} (\bibinfo {year}
  {2016})}\BibitemShut {NoStop}%
\bibitem [{\citenamefont {Hsieh}\ and\ \citenamefont
  {Hal{\'a}sz}(2017)}]{hsieh2017fractons}%
  \BibitemOpen
  \bibfield  {author} {\bibinfo {author} {\bibfnamefont {T.~H.}\ \bibnamefont
  {Hsieh}}\ and\ \bibinfo {author} {\bibfnamefont {G.~B.}\ \bibnamefont
  {Hal{\'a}sz}},\ }\href@noop {} {\bibfield  {journal} {\bibinfo  {journal}
  {\prb}\ }\textbf {\bibinfo {volume} {96}},\ \bibinfo {pages} {165105}
  (\bibinfo {year} {2017})}\BibitemShut {NoStop}%
\bibitem [{\citenamefont {Slagle}\ and\ \citenamefont
  {Kim}(2017{\natexlab{a}})}]{Slagle2017-ne}%
  \BibitemOpen
  \bibfield  {author} {\bibinfo {author} {\bibfnamefont {K.}~\bibnamefont
  {Slagle}}\ and\ \bibinfo {author} {\bibfnamefont {Y.~B.}\ \bibnamefont
  {Kim}},\ }\href@noop {} {\bibfield  {journal} {\bibinfo  {journal} {Phys.
  Rev. B}\ }\textbf {\bibinfo {volume} {96}},\ \bibinfo {pages} {165106}
  (\bibinfo {year} {2017}{\natexlab{a}})}\BibitemShut {NoStop}%
\bibitem [{\citenamefont {Hal\'asz}\ \emph {et~al.}(2017)\citenamefont
  {Hal\'asz}, \citenamefont {Hsieh},\ and\ \citenamefont
  {Balents}}]{Halasz2017-ov}%
  \BibitemOpen
  \bibfield  {author} {\bibinfo {author} {\bibfnamefont {G.~B.}\ \bibnamefont
  {Hal\'asz}}, \bibinfo {author} {\bibfnamefont {T.~H.}\ \bibnamefont {Hsieh}},
  \ and\ \bibinfo {author} {\bibfnamefont {L.}~\bibnamefont {Balents}},\
  }\href@noop {} {\bibfield  {journal} {\bibinfo  {journal} {Phys. Rev. Lett.}\
  }\textbf {\bibinfo {volume} {119}},\ \bibinfo {pages} {257202} (\bibinfo
  {year} {2017})}\BibitemShut {NoStop}%
\bibitem [{\citenamefont {Hsieh}\ and\ \citenamefont
  {Hal\'asz}(2017)}]{Hsieh2017-sc}%
  \BibitemOpen
  \bibfield  {author} {\bibinfo {author} {\bibfnamefont {T.~H.}\ \bibnamefont
  {Hsieh}}\ and\ \bibinfo {author} {\bibfnamefont {G.~B.}\ \bibnamefont
  {Hal\'asz}},\ }\href@noop {} {\bibfield  {journal} {\bibinfo  {journal}
  {Phys. Rev. B}\ }\textbf {\bibinfo {volume} {96}},\ \bibinfo {pages} {165105}
  (\bibinfo {year} {2017})}\BibitemShut {NoStop}%
\bibitem [{\citenamefont {Chamon}(2005)}]{Chamon2005-fc}%
  \BibitemOpen
  \bibfield  {author} {\bibinfo {author} {\bibfnamefont {C.}~\bibnamefont
  {Chamon}},\ }\href@noop {} {\bibfield  {journal} {\bibinfo  {journal} {Phys.
  Rev. Lett.}\ }\textbf {\bibinfo {volume} {94}},\ \bibinfo {pages} {040402}
  (\bibinfo {year} {2005})}\BibitemShut {NoStop}%
\bibitem [{\citenamefont {Ma}\ \emph {et~al.}(2017)\citenamefont {Ma},
  \citenamefont {Lake}, \citenamefont {Chen},\ and\ \citenamefont
  {Hermele}}]{Ma2017-qq}%
  \BibitemOpen
  \bibfield  {author} {\bibinfo {author} {\bibfnamefont {H.}~\bibnamefont
  {Ma}}, \bibinfo {author} {\bibfnamefont {E.}~\bibnamefont {Lake}}, \bibinfo
  {author} {\bibfnamefont {X.}~\bibnamefont {Chen}}, \ and\ \bibinfo {author}
  {\bibfnamefont {M.}~\bibnamefont {Hermele}},\ }\href@noop {} {\bibfield
  {journal} {\bibinfo  {journal} {Phys. Rev. B}\ }\textbf {\bibinfo {volume}
  {95}},\ \bibinfo {pages} {245126} (\bibinfo {year} {2017})}\BibitemShut
  {NoStop}%
\bibitem [{\citenamefont {Vijay}(2017)}]{Vijay2017-ey}%
  \BibitemOpen
  \bibfield  {author} {\bibinfo {author} {\bibfnamefont {S.}~\bibnamefont
  {Vijay}},\ }\href@noop {} {\bibfield  {journal} {\bibinfo  {journal}
  {arXiv:1701.00762}\ } (\bibinfo {year} {2017})}\BibitemShut {NoStop}%
\bibitem [{\citenamefont {Slagle}\ and\ \citenamefont
  {Kim}(2017{\natexlab{b}})}]{Slagle2017-gk}%
  \BibitemOpen
  \bibfield  {author} {\bibinfo {author} {\bibfnamefont {K.}~\bibnamefont
  {Slagle}}\ and\ \bibinfo {author} {\bibfnamefont {Y.~B.}\ \bibnamefont
  {Kim}},\ }\href@noop {} {\bibfield  {journal} {\bibinfo  {journal} {Phys.
  Rev. B}\ }\textbf {\bibinfo {volume} {96}},\ \bibinfo {pages} {195139}
  (\bibinfo {year} {2017}{\natexlab{b}})}\BibitemShut {NoStop}%
\bibitem [{\citenamefont {Ma}\ \emph {et~al.}(2018{\natexlab{a}})\citenamefont
  {Ma}, \citenamefont {Schmitz}, \citenamefont {Parameswaran}, \citenamefont
  {Hermele},\ and\ \citenamefont {Nandkishore}}]{Ma2017-cb}%
  \BibitemOpen
  \bibfield  {author} {\bibinfo {author} {\bibfnamefont {H.}~\bibnamefont
  {Ma}}, \bibinfo {author} {\bibfnamefont {A.~T.}\ \bibnamefont {Schmitz}},
  \bibinfo {author} {\bibfnamefont {S.~A.}\ \bibnamefont {Parameswaran}},
  \bibinfo {author} {\bibfnamefont {M.}~\bibnamefont {Hermele}}, \ and\
  \bibinfo {author} {\bibfnamefont {R.~M.}\ \bibnamefont {Nandkishore}},\
  }\href@noop {} {\bibfield  {journal} {\bibinfo  {journal} {Phys. Rev. B}\
  }\textbf {\bibinfo {volume} {97}},\ \bibinfo {pages} {125101} (\bibinfo
  {year} {2018}{\natexlab{a}})}\BibitemShut {NoStop}%
\bibitem [{\citenamefont {Yoshida}(2013{\natexlab{a}})}]{yoshida2013exotic}%
  \BibitemOpen
  \bibfield  {author} {\bibinfo {author} {\bibfnamefont {B.}~\bibnamefont
  {Yoshida}},\ }\href@noop {} {\bibfield  {journal} {\bibinfo  {journal}
  {\prb}\ }\textbf {\bibinfo {volume} {88}},\ \bibinfo {pages} {125122}
  (\bibinfo {year} {2013}{\natexlab{a}})}\BibitemShut {NoStop}%
\bibitem [{\citenamefont {Haah}(2011)}]{Haah2011-ny}%
  \BibitemOpen
  \bibfield  {author} {\bibinfo {author} {\bibfnamefont {J.}~\bibnamefont
  {Haah}},\ }\href@noop {} {\bibfield  {journal} {\bibinfo  {journal} {Phys.
  Rev. A}\ }\textbf {\bibinfo {volume} {83}},\ \bibinfo {pages} {042330}
  (\bibinfo {year} {2011})}\BibitemShut {NoStop}%
\bibitem [{\citenamefont {Slagle}\ and\ \citenamefont
  {Kim}(2018)}]{Slagle2017-la}%
  \BibitemOpen
  \bibfield  {author} {\bibinfo {author} {\bibfnamefont {K.}~\bibnamefont
  {Slagle}}\ and\ \bibinfo {author} {\bibfnamefont {Y.~B.}\ \bibnamefont
  {Kim}},\ }\href@noop {} {\bibfield  {journal} {\bibinfo  {journal} {Phys.
  Rev. B}\ }\textbf {\bibinfo {volume} {97}},\ \bibinfo {pages} {165106}
  (\bibinfo {year} {2018})}\BibitemShut {NoStop}%
\bibitem [{\citenamefont {Shirley}\ \emph {et~al.}(2017)\citenamefont
  {Shirley}, \citenamefont {Slagle}, \citenamefont {Wang},\ and\ \citenamefont
  {Chen}}]{shirley2017fracton}%
  \BibitemOpen
  \bibfield  {author} {\bibinfo {author} {\bibfnamefont {W.}~\bibnamefont
  {Shirley}}, \bibinfo {author} {\bibfnamefont {K.}~\bibnamefont {Slagle}},
  \bibinfo {author} {\bibfnamefont {Z.}~\bibnamefont {Wang}}, \ and\ \bibinfo
  {author} {\bibfnamefont {X.}~\bibnamefont {Chen}},\ }\href@noop {} {\bibfield
   {journal} {\bibinfo  {journal} {arXiv:1712.05892}\ } (\bibinfo {year}
  {2017})}\BibitemShut {NoStop}%
\bibitem [{\citenamefont {Pretko}\ and\ \citenamefont
  {Radzihovsky}(2017)}]{pretko2017fracton}%
  \BibitemOpen
  \bibfield  {author} {\bibinfo {author} {\bibfnamefont {M.}~\bibnamefont
  {Pretko}}\ and\ \bibinfo {author} {\bibfnamefont {L.}~\bibnamefont
  {Radzihovsky}},\ }\href@noop {} {\bibfield  {journal} {\bibinfo  {journal}
  {arXiv:1711.11044}\ } (\bibinfo {year} {2017})}\BibitemShut {NoStop}%
\bibitem [{\citenamefont {Ma}\ \emph {et~al.}(2018{\natexlab{b}})\citenamefont
  {Ma}, \citenamefont {Hermele},\ and\ \citenamefont {Chen}}]{ma2018fracton}%
  \BibitemOpen
  \bibfield  {author} {\bibinfo {author} {\bibfnamefont {H.}~\bibnamefont
  {Ma}}, \bibinfo {author} {\bibfnamefont {M.}~\bibnamefont {Hermele}}, \ and\
  \bibinfo {author} {\bibfnamefont {X.}~\bibnamefont {Chen}},\ }\href@noop {}
  {\bibfield  {journal} {\bibinfo  {journal} {arXiv:1802.10108}\ } (\bibinfo
  {year} {2018}{\natexlab{b}})}\BibitemShut {NoStop}%
\bibitem [{\citenamefont {Prem}\ \emph
  {et~al.}(2017{\natexlab{a}})\citenamefont {Prem}, \citenamefont {Pretko},\
  and\ \citenamefont {Nandkishore}}]{prem2017emergent}%
  \BibitemOpen
  \bibfield  {author} {\bibinfo {author} {\bibfnamefont {A.}~\bibnamefont
  {Prem}}, \bibinfo {author} {\bibfnamefont {M.}~\bibnamefont {Pretko}}, \ and\
  \bibinfo {author} {\bibfnamefont {R.}~\bibnamefont {Nandkishore}},\
  }\href@noop {} {\bibfield  {journal} {\bibinfo  {journal} {arXiv:1709.09673}\
  } (\bibinfo {year} {2017}{\natexlab{a}})}\BibitemShut {NoStop}%
\bibitem [{\citenamefont {Pretko}(2017)}]{pretko2017subdimensional}%
  \BibitemOpen
  \bibfield  {author} {\bibinfo {author} {\bibfnamefont {M.}~\bibnamefont
  {Pretko}},\ }\href@noop {} {\bibfield  {journal} {\bibinfo  {journal} {\prb}\
  }\textbf {\bibinfo {volume} {95}},\ \bibinfo {pages} {115139} (\bibinfo
  {year} {2017})}\BibitemShut {NoStop}%
\bibitem [{\citenamefont {Bulmash}\ and\ \citenamefont
  {Barkeshli}(2018)}]{bulmash2018higgs}%
  \BibitemOpen
  \bibfield  {author} {\bibinfo {author} {\bibfnamefont {D.}~\bibnamefont
  {Bulmash}}\ and\ \bibinfo {author} {\bibfnamefont {M.}~\bibnamefont
  {Barkeshli}},\ }\href@noop {} {\bibfield  {journal} {\bibinfo  {journal}
  {arXiv:1802.10099}\ } (\bibinfo {year} {2018})}\BibitemShut {NoStop}%
\bibitem [{\citenamefont {Prem}\ \emph
  {et~al.}(2017{\natexlab{b}})\citenamefont {Prem}, \citenamefont {Haah},\ and\
  \citenamefont {Nandkishore}}]{Prem2017-ql}%
  \BibitemOpen
  \bibfield  {author} {\bibinfo {author} {\bibfnamefont {A.}~\bibnamefont
  {Prem}}, \bibinfo {author} {\bibfnamefont {J.}~\bibnamefont {Haah}}, \ and\
  \bibinfo {author} {\bibfnamefont {R.}~\bibnamefont {Nandkishore}},\
  }\href@noop {} {\bibfield  {journal} {\bibinfo  {journal} {Phys. Rev. B}\
  }\textbf {\bibinfo {volume} {95}},\ \bibinfo {pages} {155133} (\bibinfo
  {year} {2017}{\natexlab{b}})}\BibitemShut {NoStop}%
\bibitem [{\citenamefont {{Bulmash}}\ and\ \citenamefont
  {{Barkeshli}}(2018)}]{2018arXiv180601855B}%
  \BibitemOpen
  \bibfield  {author} {\bibinfo {author} {\bibfnamefont {D.}~\bibnamefont
  {{Bulmash}}}\ and\ \bibinfo {author} {\bibfnamefont {M.}~\bibnamefont
  {{Barkeshli}}},\ }\href@noop {} {\bibfield  {journal} {\bibinfo  {journal}
  {arXiv:1806.01855}\ } (\bibinfo {year} {2018})}\BibitemShut {NoStop}%
\bibitem [{\citenamefont {You}\ \emph {et~al.}(2018{\natexlab{b}})\citenamefont
  {You}, \citenamefont {Devakul}, \citenamefont {Burnell},\ and\ \citenamefont
  {Sondhi}}]{you2018subsystem}%
  \BibitemOpen
  \bibfield  {author} {\bibinfo {author} {\bibfnamefont {Y.}~\bibnamefont
  {You}}, \bibinfo {author} {\bibfnamefont {T.}~\bibnamefont {Devakul}},
  \bibinfo {author} {\bibfnamefont {F.}~\bibnamefont {Burnell}}, \ and\
  \bibinfo {author} {\bibfnamefont {S.}~\bibnamefont {Sondhi}},\ }\href@noop {}
  {\bibfield  {journal} {\bibinfo  {journal} {\prb}\ }\textbf {\bibinfo
  {volume} {98}},\ \bibinfo {pages} {035112} (\bibinfo {year}
  {2018}{\natexlab{b}})}\BibitemShut {NoStop}%
\bibitem [{\citenamefont {Devakul}\ \emph
  {et~al.}(2018{\natexlab{a}})\citenamefont {Devakul}, \citenamefont {You},
  \citenamefont {Burnell},\ and\ \citenamefont {Sondhi}}]{devakul2018fractal}%
  \BibitemOpen
  \bibfield  {author} {\bibinfo {author} {\bibfnamefont {T.}~\bibnamefont
  {Devakul}}, \bibinfo {author} {\bibfnamefont {Y.}~\bibnamefont {You}},
  \bibinfo {author} {\bibfnamefont {F.}~\bibnamefont {Burnell}}, \ and\
  \bibinfo {author} {\bibfnamefont {S.}~\bibnamefont {Sondhi}},\ }\href@noop {}
  {\bibfield  {journal} {\bibinfo  {journal} {arXiv:1805.04097}\ } (\bibinfo
  {year} {2018}{\natexlab{a}})}\BibitemShut {NoStop}%
\bibitem [{\citenamefont {You}\ \emph {et~al.}(2018{\natexlab{c}})\citenamefont
  {You}, \citenamefont {Devakul}, \citenamefont {Burnell},\ and\ \citenamefont
  {Sondhi}}]{you2018symmetric}%
  \BibitemOpen
  \bibfield  {author} {\bibinfo {author} {\bibfnamefont {Y.}~\bibnamefont
  {You}}, \bibinfo {author} {\bibfnamefont {T.}~\bibnamefont {Devakul}},
  \bibinfo {author} {\bibfnamefont {F.}~\bibnamefont {Burnell}}, \ and\
  \bibinfo {author} {\bibfnamefont {S.}~\bibnamefont {Sondhi}},\ }\href@noop {}
  {\bibfield  {journal} {\bibinfo  {journal} {arXiv:1805.09800}\ } (\bibinfo
  {year} {2018}{\natexlab{c}})}\BibitemShut {NoStop}%
  \bibitem [{\citenamefont {{Yan}}(2018)}]{2018arXiv180705942}%
  \BibitemOpen
  \bibfield  {author} {\bibinfo {author} {\bibfnamefont {H.}~\bibnamefont
  {{Yan}}},\ }\href@noop {} {\bibfield  {journal} {\bibinfo  {journal}
  {arXiv:1807.05942}\ } (\bibinfo {year} {2018})}\BibitemShut {NoStop}%
\bibitem [{\citenamefont {Slagle}\ and\ \citenamefont
  {Kim}(2017{\natexlab{c}})}]{slagle2017fracton}%
  \BibitemOpen
  \bibfield  {author} {\bibinfo {author} {\bibfnamefont {K.}~\bibnamefont
  {Slagle}}\ and\ \bibinfo {author} {\bibfnamefont {Y.~B.}\ \bibnamefont
  {Kim}},\ }\href@noop {} {\bibfield  {journal} {\bibinfo  {journal} {\prb}\
  }\textbf {\bibinfo {volume} {96}},\ \bibinfo {pages} {165106} (\bibinfo
  {year} {2017}{\natexlab{c}})}\BibitemShut {NoStop}%
\bibitem [{\citenamefont {Landau}\ \emph {et~al.}(2016)\citenamefont {Landau},
  \citenamefont {Plugge}, \citenamefont {Sela}, \citenamefont {Altland},
  \citenamefont {Albrecht},\ and\ \citenamefont {Egger}}]{landau2016towards}%
  \BibitemOpen
  \bibfield  {author} {\bibinfo {author} {\bibfnamefont {L.}~\bibnamefont
  {Landau}}, \bibinfo {author} {\bibfnamefont {S.}~\bibnamefont {Plugge}},
  \bibinfo {author} {\bibfnamefont {E.}~\bibnamefont {Sela}}, \bibinfo {author}
  {\bibfnamefont {A.}~\bibnamefont {Altland}}, \bibinfo {author} {\bibfnamefont
  {S.}~\bibnamefont {Albrecht}}, \ and\ \bibinfo {author} {\bibfnamefont
  {R.}~\bibnamefont {Egger}},\ }\href@noop {} {\bibfield  {journal} {\bibinfo
  {journal} {\prl}\ }\textbf {\bibinfo {volume} {116}},\ \bibinfo {pages}
  {050501} (\bibinfo {year} {2016})}\BibitemShut {NoStop}%
\bibitem [{SM()}]{SM}%
  \BibitemOpen
  \href@noop {} {\bibinfo  {journal} {Supplementary Material}\ }\BibitemShut
  {NoStop}%
\bibitem [{Note1()}]{Note1}%
  \BibitemOpen
\bibfield  {journal} {  }\bibinfo {note} {Unlike the corner mode in HOTSC, this
  $\epsilon $ fermion zero mode does not carry fermion parity.}\BibitemShut
  {Stop}%
\bibitem [{\citenamefont {Bomb{\'\i}n}(2010)}]{bombin2010topological}%
  \BibitemOpen
  \bibfield  {author} {\bibinfo {author} {\bibfnamefont {H.}~\bibnamefont
  {Bomb{\'\i}n}},\ }\href@noop {} {\bibfield  {journal} {\bibinfo  {journal}
  {\prl}\ }\textbf {\bibinfo {volume} {105}},\ \bibinfo {pages} {030403}
  (\bibinfo {year} {2010})}\BibitemShut {NoStop}%
\bibitem [{\citenamefont {Brown}\ \emph {et~al.}(2017)\citenamefont {Brown},
  \citenamefont {Laubscher}, \citenamefont {Kesselring},\ and\ \citenamefont
  {Wootton}}]{brown2017}%
  \BibitemOpen
  \bibfield  {author} {\bibinfo {author} {\bibfnamefont {B.~J.}\ \bibnamefont
  {Brown}}, \bibinfo {author} {\bibfnamefont {K.}~\bibnamefont {Laubscher}},
  \bibinfo {author} {\bibfnamefont {M.~S.}\ \bibnamefont {Kesselring}}, \ and\
  \bibinfo {author} {\bibfnamefont {J.~R.}\ \bibnamefont {Wootton}},\
  }\href@noop {} {\bibfield  {journal} {\bibinfo  {journal} {Phys. Rev. X}\
  }\textbf {\bibinfo {volume} {7}},\ \bibinfo {pages} {021029} (\bibinfo {year}
  {2017})}\BibitemShut {NoStop}%
\bibitem [{\citenamefont {Thorngren}\ and\ \citenamefont
  {Else}(2018)}]{thorngren2018gauging}%
  \BibitemOpen
  \bibfield  {author} {\bibinfo {author} {\bibfnamefont {R.}~\bibnamefont
  {Thorngren}}\ and\ \bibinfo {author} {\bibfnamefont {D.~V.}\ \bibnamefont
  {Else}},\ }\href@noop {} {\bibfield  {journal} {\bibinfo  {journal} {Phys.\
  Rev.\ X}\ }\textbf {\bibinfo {volume} {8}},\ \bibinfo {pages} {011040}
  (\bibinfo {year} {2018})}\BibitemShut {NoStop}%
\bibitem [{\citenamefont {{Else}}\ and\ \citenamefont
  {{Thorngren}}(2018)}]{else2018}%
  \BibitemOpen
  \bibfield  {author} {\bibinfo {author} {\bibfnamefont {D.~V.}\ \bibnamefont
  {{Else}}}\ and\ \bibinfo {author} {\bibfnamefont {R.}~\bibnamefont
  {{Thorngren}}},\ }\href@noop {} {\bibfield  {journal} {\bibinfo  {journal}
  {arXiv:1810.10539}\ } (\bibinfo {year} {2018})}\BibitemShut {NoStop}%
\bibitem [{\citenamefont {Barkeshli}\ \emph {et~al.}(2014)\citenamefont
  {Barkeshli}, \citenamefont {Bonderson}, \citenamefont {Cheng},\ and\
  \citenamefont {Wang}}]{barkeshli2014symmetry}%
  \BibitemOpen
  \bibfield  {author} {\bibinfo {author} {\bibfnamefont {M.}~\bibnamefont
  {Barkeshli}}, \bibinfo {author} {\bibfnamefont {P.}~\bibnamefont
  {Bonderson}}, \bibinfo {author} {\bibfnamefont {M.}~\bibnamefont {Cheng}}, \
  and\ \bibinfo {author} {\bibfnamefont {Z.}~\bibnamefont {Wang}},\ }\href@noop
  {} {\bibfield  {journal} {\bibinfo  {journal} {arXiv:1410.4540}\ } (\bibinfo
  {year} {2014})}\BibitemShut {NoStop}%
\bibitem [{\citenamefont {You}\ and\ \citenamefont {von Oppen}()}]{unpub}%
  \BibitemOpen
  \bibfield  {author} {\bibinfo {author} {\bibfnamefont {Y.}~\bibnamefont
  {You}}\ and\ \bibinfo {author} {\bibfnamefont {F.}~\bibnamefont {von
  Oppen}},\ }\href@noop {} {\bibinfo  {journal} {unpublished}\ }\BibitemShut
  {NoStop}%
\bibitem [{\citenamefont {Devakul}\ \emph
  {et~al.}(2018{\natexlab{b}})\citenamefont {Devakul}, \citenamefont
  {Williamson},\ and\ \citenamefont {You}}]{devakul2018strong}%
  \BibitemOpen
\bibfield  {journal} {  }\bibfield  {author} {\bibinfo {author} {\bibfnamefont
  {T.}~\bibnamefont {Devakul}}, \bibinfo {author} {\bibfnamefont {D.~J.}\
  \bibnamefont {Williamson}}, \ and\ \bibinfo {author} {\bibfnamefont
  {Y.}~\bibnamefont {You}},\ }\href@noop {} {\bibfield  {journal} {\bibinfo
  {journal} {arXiv:1808.05300}\ } (\bibinfo {year}
  {2018}{\natexlab{b}})}\BibitemShut {NoStop}%
\bibitem [{\citenamefont {Vijay}\ \emph
  {et~al.}(2015{\natexlab{a}})\citenamefont {Vijay}, \citenamefont {Haah},\
  and\ \citenamefont {Fu}}]{Vijay2015-jj}%
  \BibitemOpen
  \bibfield  {author} {\bibinfo {author} {\bibfnamefont {S.}~\bibnamefont
  {Vijay}}, \bibinfo {author} {\bibfnamefont {J.}~\bibnamefont {Haah}}, \ and\
  \bibinfo {author} {\bibfnamefont {L.}~\bibnamefont {Fu}},\ }\href@noop {}
  {\bibfield  {journal} {\bibinfo  {journal} {Phys. Rev. B}\ }\textbf {\bibinfo
  {volume} {92}},\ \bibinfo {pages} {235136} (\bibinfo {year}
  {2015}{\natexlab{a}})}\BibitemShut {NoStop}%
\bibitem [{\citenamefont {Vijay}\ \emph
  {et~al.}(2015{\natexlab{b}})\citenamefont {Vijay}, \citenamefont {Hsieh},\
  and\ \citenamefont {Fu}}]{vijay2015majorana}%
  \BibitemOpen
  \bibfield  {author} {\bibinfo {author} {\bibfnamefont {S.}~\bibnamefont
  {Vijay}}, \bibinfo {author} {\bibfnamefont {T.~H.}\ \bibnamefont {Hsieh}}, \
  and\ \bibinfo {author} {\bibfnamefont {L.}~\bibnamefont {Fu}},\ }\href@noop
  {} {\bibfield  {journal} {\bibinfo  {journal} {Phys.\ Rev.\ X}\ }\textbf
  {\bibinfo {volume} {5}},\ \bibinfo {pages} {041038} (\bibinfo {year}
  {2015}{\natexlab{b}})}\BibitemShut {NoStop}%
\bibitem [{\citenamefont {Karzig}\ \emph {et~al.}(2017)\citenamefont {Karzig},
  \citenamefont {Knapp}, \citenamefont {Lutchyn}, \citenamefont {Bonderson},
  \citenamefont {Hastings}, \citenamefont {Nayak}, \citenamefont {Alicea},
  \citenamefont {Flensberg}, \citenamefont {Plugge}, \citenamefont {Oreg} \emph
  {et~al.}}]{karzig2017scalable}%
  \BibitemOpen
  \bibfield  {author} {\bibinfo {author} {\bibfnamefont {T.}~\bibnamefont
  {Karzig}}, \bibinfo {author} {\bibfnamefont {C.}~\bibnamefont {Knapp}},
  \bibinfo {author} {\bibfnamefont {R.~M.}\ \bibnamefont {Lutchyn}}, \bibinfo
  {author} {\bibfnamefont {P.}~\bibnamefont {Bonderson}}, \bibinfo {author}
  {\bibfnamefont {M.~B.}\ \bibnamefont {Hastings}}, \bibinfo {author}
  {\bibfnamefont {C.}~\bibnamefont {Nayak}}, \bibinfo {author} {\bibfnamefont
  {J.}~\bibnamefont {Alicea}}, \bibinfo {author} {\bibfnamefont
  {K.}~\bibnamefont {Flensberg}}, \bibinfo {author} {\bibfnamefont
  {S.}~\bibnamefont {Plugge}}, \bibinfo {author} {\bibfnamefont
  {Y.}~\bibnamefont {Oreg}},  \emph {et~al.},\ }\href@noop {} {\bibfield
  {journal} {\bibinfo  {journal} {\prb}\ }\textbf {\bibinfo {volume} {95}},\
  \bibinfo {pages} {235305} (\bibinfo {year} {2017})}\BibitemShut {NoStop}%
\bibitem [{\citenamefont {{Wille}}\ \emph {et~al.}(2018)\citenamefont
  {{Wille}}, \citenamefont {{Egger}}, \citenamefont {{Eisert}},\ and\
  \citenamefont {{Altland}}}]{2018arXiv180804529W}%
  \BibitemOpen
  \bibfield  {author} {\bibinfo {author} {\bibfnamefont {C.}~\bibnamefont
  {{Wille}}}, \bibinfo {author} {\bibfnamefont {R.}~\bibnamefont {{Egger}}},
  \bibinfo {author} {\bibfnamefont {J.}~\bibnamefont {{Eisert}}}, \ and\
  \bibinfo {author} {\bibfnamefont {A.}~\bibnamefont {{Altland}}},\ }\href@noop
  {} {\bibfield  {journal} {\bibinfo  {journal} {arXiv:1808.04529}\ } (\bibinfo
  {year} {2018})}\BibitemShut {NoStop}%
\bibitem [{\citenamefont {{Thomson}}\ and\ \citenamefont
  {{Pientka}}(2018)}]{2018arXiv180709291T}%
  \BibitemOpen
  \bibfield  {author} {\bibinfo {author} {\bibfnamefont {A.}~\bibnamefont
  {{Thomson}}}\ and\ \bibinfo {author} {\bibfnamefont {F.}~\bibnamefont
  {{Pientka}}},\ }\href@noop {} {\bibfield  {journal} {\bibinfo  {journal}
  {arXiv:1807.09291}\ } (\bibinfo {year} {2018})}\BibitemShut {NoStop}%
\bibitem [{\citenamefont {{Sagi}}\ \emph {et~al.}(2018)\citenamefont {{Sagi}},
  \citenamefont {{Ebisu}}, \citenamefont {{Tanaka}}, \citenamefont {{Stern}},\
  and\ \citenamefont {{Oreg}}}]{2018arXiv180603304S}%
  \BibitemOpen
  \bibfield  {author} {\bibinfo {author} {\bibfnamefont {E.}~\bibnamefont
  {{Sagi}}}, \bibinfo {author} {\bibfnamefont {H.}~\bibnamefont {{Ebisu}}},
  \bibinfo {author} {\bibfnamefont {Y.}~\bibnamefont {{Tanaka}}}, \bibinfo
  {author} {\bibfnamefont {A.}~\bibnamefont {{Stern}}}, \ and\ \bibinfo
  {author} {\bibfnamefont {Y.}~\bibnamefont {{Oreg}}},\ }\href@noop {}
  {\bibfield  {journal} {\bibinfo  {journal} {arXiv:1806.03304}\ } (\bibinfo
  {year} {2018})}\BibitemShut {NoStop}%
\bibitem [{\citenamefont {You}\ \emph {et~al.}(2013)\citenamefont {You},
  \citenamefont {Jian},\ and\ \citenamefont {Wen}}]{you2012synthetic}%
  \BibitemOpen
  \bibfield  {author} {\bibinfo {author} {\bibfnamefont {Y.-Z.}\ \bibnamefont
  {You}}, \bibinfo {author} {\bibfnamefont {C.-M.}\ \bibnamefont {Jian}}, \
  and\ \bibinfo {author} {\bibfnamefont {X.-G.}\ \bibnamefont {Wen}},\
  }\href@noop {} {\bibfield  {journal} {\bibinfo  {journal} {\prb}\ }\textbf
  {\bibinfo {volume} {87}},\ \bibinfo {pages} {045106} (\bibinfo {year}
  {2013})}\BibitemShut {NoStop}%
\bibitem [{\citenamefont {Barkeshli}\ \emph {et~al.}(2013)\citenamefont
  {Barkeshli}, \citenamefont {Jian},\ and\ \citenamefont
  {Qi}}]{barkeshli2013classification}%
  \BibitemOpen
  \bibfield  {author} {\bibinfo {author} {\bibfnamefont {M.}~\bibnamefont
  {Barkeshli}}, \bibinfo {author} {\bibfnamefont {C.-M.}\ \bibnamefont {Jian}},
  \ and\ \bibinfo {author} {\bibfnamefont {X.-L.}\ \bibnamefont {Qi}},\
  }\href@noop {} {\bibfield  {journal} {\bibinfo  {journal} {\prb}\ }\textbf
  {\bibinfo {volume} {88}},\ \bibinfo {pages} {241103} (\bibinfo {year}
  {2013})}\BibitemShut {NoStop}%
\bibitem [{\citenamefont {Chen}\ \emph {et~al.}(2010)\citenamefont {Chen},
  \citenamefont {Gu},\ and\ \citenamefont {Wen}}]{chen2010local}%
  \BibitemOpen
  \bibfield  {author} {\bibinfo {author} {\bibfnamefont {X.}~\bibnamefont
  {Chen}}, \bibinfo {author} {\bibfnamefont {Z.-C.}\ \bibnamefont {Gu}}, \ and\
  \bibinfo {author} {\bibfnamefont {X.-G.}\ \bibnamefont {Wen}},\ }\href@noop
  {} {\bibfield  {journal} {\bibinfo  {journal} {\prb}\ }\textbf {\bibinfo
  {volume} {82}},\ \bibinfo {pages} {155138} (\bibinfo {year}
  {2010})}\BibitemShut {NoStop}%
\bibitem [{\citenamefont {Levin}(2013)}]{levin2013protected}%
  \BibitemOpen
  \bibfield  {author} {\bibinfo {author} {\bibfnamefont {M.}~\bibnamefont
  {Levin}},\ }\href@noop {} {\bibfield  {journal} {\bibinfo  {journal} {Phys.\
  Rev.\ X}\ }\textbf {\bibinfo {volume} {3}},\ \bibinfo {pages} {021009}
  (\bibinfo {year} {2013})}\BibitemShut {NoStop}%
\bibitem [{\citenamefont {Litinski}\ and\ \citenamefont {von
  Oppen}(2018)}]{litinski2018quantum}%
  \BibitemOpen
  \bibfield  {author} {\bibinfo {author} {\bibfnamefont {D.}~\bibnamefont
  {Litinski}}\ and\ \bibinfo {author} {\bibfnamefont {F.}~\bibnamefont {von
  Oppen}},\ }\href@noop {} {\bibfield  {journal} {\bibinfo  {journal} {\prb}\
  }\textbf {\bibinfo {volume} {97}},\ \bibinfo {pages} {205404} (\bibinfo
  {year} {2018})}\BibitemShut {NoStop}%
\bibitem [{\citenamefont {Bombin}\ and\ \citenamefont
  {Martin-Delgado}(2006)}]{bombin2006topological}%
  \BibitemOpen
  \bibfield  {author} {\bibinfo {author} {\bibfnamefont {H.}~\bibnamefont
  {Bombin}}\ and\ \bibinfo {author} {\bibfnamefont {M.~A.}\ \bibnamefont
  {Martin-Delgado}},\ }\href@noop {} {\bibfield  {journal} {\bibinfo  {journal}
  {\prl}\ }\textbf {\bibinfo {volume} {97}},\ \bibinfo {pages} {180501}
  (\bibinfo {year} {2006})}\BibitemShut {NoStop}%
\bibitem [{\citenamefont {Landahl}\ \emph {et~al.}(2011)\citenamefont
  {Landahl}, \citenamefont {Anderson},\ and\ \citenamefont
  {Rice}}]{landahl2011fault}%
  \BibitemOpen
  \bibfield  {author} {\bibinfo {author} {\bibfnamefont {A.~J.}\ \bibnamefont
  {Landahl}}, \bibinfo {author} {\bibfnamefont {J.~T.}\ \bibnamefont
  {Anderson}}, \ and\ \bibinfo {author} {\bibfnamefont {P.~R.}\ \bibnamefont
  {Rice}},\ }\href@noop {} {\bibfield  {journal} {\bibinfo  {journal}
  {arXiv:1108.5738}\ } (\bibinfo {year} {2011})}\BibitemShut {NoStop}%
\bibitem [{\citenamefont {Yoshida}(2013{\natexlab{b}})}]{Yoshida2013-of}%
  \BibitemOpen
  \bibfield  {author} {\bibinfo {author} {\bibfnamefont {B.}~\bibnamefont
  {Yoshida}},\ }\href@noop {} {\bibfield  {journal} {\bibinfo  {journal} {Phys.
  Rev. B}\ }\textbf {\bibinfo {volume} {88}},\ \bibinfo {pages} {125122}
  (\bibinfo {year} {2013}{\natexlab{b}})}\BibitemShut {NoStop}%
\end{thebibliography}
\end{document}